\documentclass[iop]{emulateapj}
\usepackage[T1]{fontenc}
\usepackage{amsmath}
\usepackage{amssymb}	
\usepackage{graphicx,aas_macros}
\usepackage{subfigure}
\usepackage[breaklinks,colorlinks,citecolor=blue]{hyperref}
\usepackage[all]{hypcap}
\usepackage{bm}
\usepackage{color}
\usepackage[dvipsnames]{xcolor}

\def\stacksymbols #1#2#3#4{\def\theguybelow{#2}
        \def\verticalposition{\lower#3pt}
        \def\spacingwithinsymbol{\baselineskip0pt\lineskip#4pt}
        \mathrel{\mathpalette\intermediary#1}}
\def\intermediary #1#2{\verticalposition\vbox{\spacingwithinsymbol
        \everycr={}\tabskip0pt
        \halign{$\mathsurround0pt#1\hfil##\hfil$\crcr#2\crcr
                \theguybelow\crcr}}}

\def\lta{\stacksymbols{<}{\sim}{2.5}{.2}}
\def\gta{\stacksymbols{>}{\sim}{2.5}{.2}}

\newcommand{\be}{\begin{equation}}
\newcommand{\ee}{\end{equation}}
\newcommand{\bea}{\begin{eqnarray}}
\newcommand{\eea}{\end{eqnarray}}

\newcommand{\msun}{M_{\odot}}

\begin{document}

\shortauthors{Gaspari \& S\k{a}dowski}
\shorttitle{Linking the micro to macro properties of AGN feedback}
\title{Unifying the micro and macro properties of AGN feeding and feedback}
\author{Massimo Gaspari$^{1,*,\dagger}$ \& Aleksander S\k{a}dowski$^{2,*}$}
\affil{$^1$Department of Astrophysical Sciences, Princeton University, 4 Ivy Lane, Princeton, NJ 08544-1001 USA; mgaspari@astro.princeton.edu\\ 
 $^2$MIT Kavli Institute for Astrophysics and Space Research, 77 Massachusetts Ave, Cambridge, MA 02139, USA; asadowsk@mit.edu}
\altaffiltext{\hspace{-0.15in} * }{\ \,Einstein Fellow.}
\altaffiltext{\hspace{-0.15in} $\dagger$ }{\ \,Spitzer Fellow.}

\begin{abstract}
\noindent
We unify the feeding and feedback of supermassive black holes with the global properties of galaxies, groups, and clusters, by linking for the first time the physical mechanical efficiency at the horizon and Mpc scale. The macro hot halo is tightly constrained by the absence of overheating and overcooling as probed by X-ray data and hydrodynamic simulations ($\varepsilon_{\rm BH} \simeq 10^{-3}\,T_{\rm x,7.4}$). The micro flow is shaped by general relativistic effects tracked by state-of-the-art GR-RMHD simulations ($\varepsilon_\bullet \simeq 0.03$). The SMBH properties are tied to the X-ray halo temperature $T_{\rm x}$, or related cosmic scaling relation (as $L_{\rm x}$). The model is minimally based on first principles, as conservation of energy and mass recycling. The inflow occurs via chaotic cold accretion (CCA), the rain of cold clouds condensing out of the quenched cooling flow and recurrently funneled via inelastic collisions. Within 100s gravitational radii, the accretion energy is transformed into ultrafast $10^4$\,km\,s$^{-1}$ outflows (UFOs) ejecting most of the inflowing mass. At larger radii the energy-driven outflow entrains progressively more mass: at roughly kpc scale, the velocities of the hot/warm/cold outflows are a few $10^3$, 1000, 500\,km\,s$^{-1}$, with median mass rates $\sim$\,10, 100, several 100\,$\msun$\,yr$^{-1}$, respectively. The unified CCA model is consistent with the observations of nuclear UFOs, and ionized, neutral, and molecular macro outflows. We provide step-by-step implementation for subgrid simulations, (semi)analytic works, or observational interpretations which require self-regulated AGN feedback at coarse scales, avoiding the a-posteriori~fine-tuning~of~efficiencies. 
\vspace{+0.22cm}
\end{abstract}


\section{Introduction} \label{s:intro}
\setcounter{footnote}{0}
\noindent
Last-decade observations and simulations have shown that 
supermassive black holes (SMBHs) and cosmic structures are not separate elements of the universe (\citealt{Heckman:2014} for a review).
While cosmic structures are characterized by virial radii\footnote{The radius $r_\delta$ encloses $\delta$ times the critical overdensity $\rho_{\rm c}(z)=3 H^2(z)/(8\pi G)$ ($H$ is the Hubble parameter; $H_0\simeq70$\,km\,s$^{-1}$\,Mpc$^{-1}$) giving an enclosed mass $M_{\delta} = (4\pi/3)\delta \rho_{\rm c}(z) r_\delta^3$; $\delta\simeq100$ for the virial radius and 500 for observational constraints.
}
$r_{100}$ ($\sim$\,Mpc), SMBHs have a characteristic Schwarzschild radius $r_{\rm S} = 2\,G M_\bullet/c^2$ ($10^{-4}$\,pc for $M_\bullet = 10^9\ \msun$), implying a difference of 10 dex in length scale. This magnitude of separation might strike as insurmountable, however, black holes would not exist without matter feeding them, and cosmic structures would tend to a quick cold death without feedback from SMBHs (often called active galactic nuclei -- AGN -- 
to emphasize such role), thus creating a symbiotic relation.

\begin{figure*}
\centering
\includegraphics[width=2.05\columnwidth]{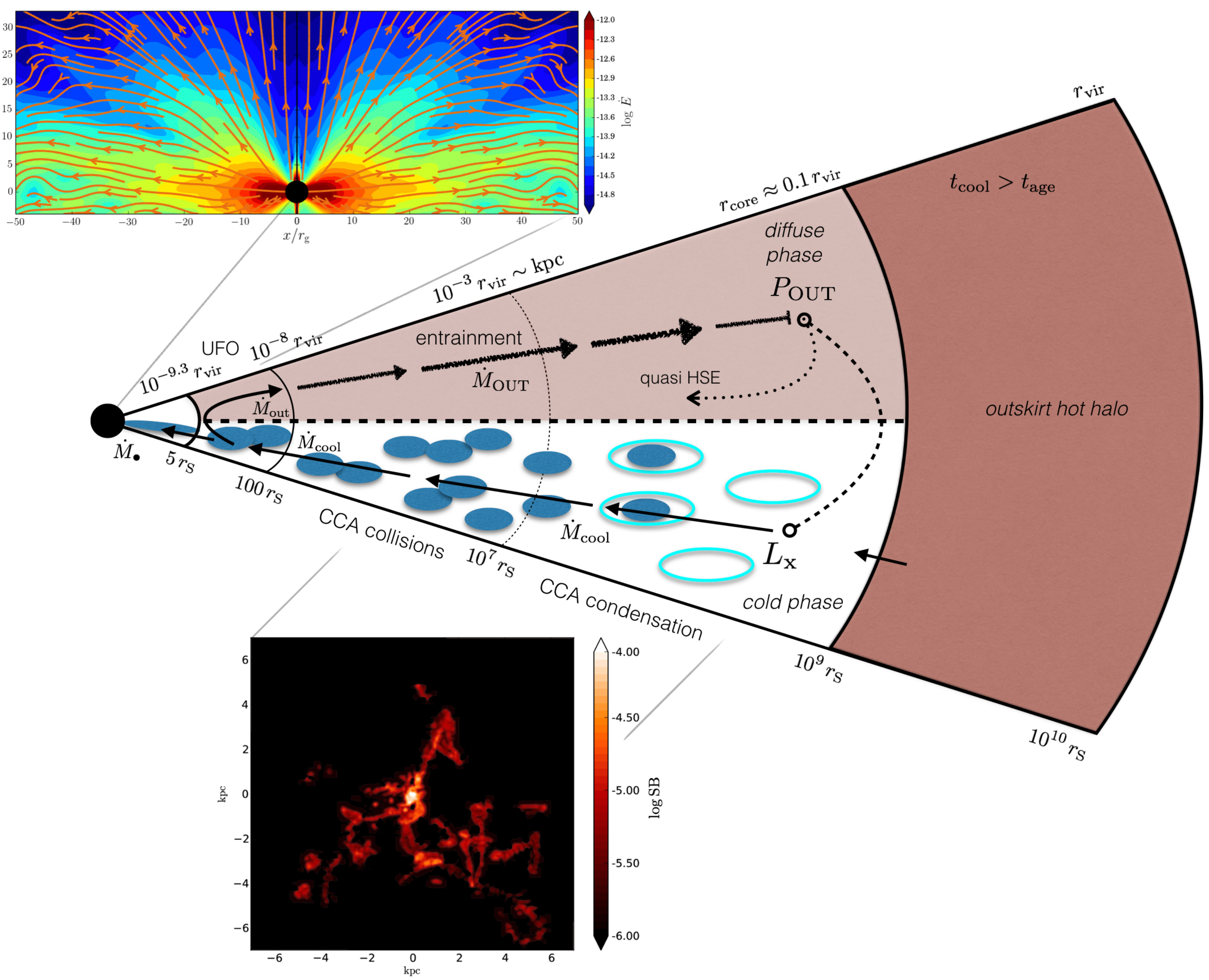}
\caption{{\it Middle} -- Diagram of the multiphase accretion inflow and outflow covering the entire range of scales, from the inner SMBH horizon to the virial radius of the galaxy, group, or cluster. The self-regulated AGN feedback loop works as follows. The turbulent gaseous halo condenses in localized, large-scale high density peaks (cyan), leading to the drop out of cold clouds and warm filaments (blue). The clouds rain down and recurrently collide in chaotic and inelastic way (CCA), canceling angular momentum and flowing toward the SMBH. The mass inflow rate originates from the {\it quenched} X-ray cooling rate within the core region. 
Within $\sim$\,$100\,r_{\rm S}$, the gravitational accretion process releases ultrafast outflows (UFOs), while only a small gas fraction is sinked through the horizon (this is balanced by a net inflow from the outskirts). 
The outflows slow down at larger radii, entraining the gas of the background profile. The energy is thermalized in the core, balancing the X-ray luminosity. The CCA rain is thus stopped, and so are the outflows, allowing the global halo to restore the quasi HSE profile. As cooling resumes without a source of heating, another cycle of CCA rain and collisions, mass ejection and entrainment, and restoration is triggered, consistently with X-ray data. The system conserves total energy and mass in a gentle recycling multiphase flow.\\
{\it Top} -- GR-RMHD simulation of the micro flow (\S\ref{s:micro}), showing the magnitude and streamlines of the total energy flux (from \citealt{Sadowski:2016_thick}; code units) which is dominated by the kinetic component with $\epsilon_\bullet \simeq 0.03$ (see SG17 for more details on the mechanical efficiency).\\
  {\it Bottom} -- Multiphase hydro simulation of the macro flow tracking the CCA evolution (from \citealt{Gaspari:2017}:  \S\ref{s:macro}). The map shows the surface brightness (erg\,s\,$^{-1}$\,cm$^{-2}$\,sr$^{-1}$) of the filamentary warm phase condensed out of the turbulent hot halo and chaotically colliding.
} 
\label{f:sketch}
\end{figure*}

At the present, no simulation is capable of covering simultaneously the 10 dex dynamic range involving SMBH feeding and feedback (Fig.~\ref{f:sketch}), and to track the evolution from 0.1 yr to 10 Gyr.
Recent attempts have been made in the direction of linking the large-scale multiphase gaseous halos of galaxies (ISM), groups (IGM), and clusters (ICM) down to the subpc accretion scale (e.g., \citealt{Gaspari:2015_cca,Gaspari:2017} -- G15, G17).
The dark matter halos heat up the diffuse gas during gravitational collapse, creating stratified hot plasma halos ($\sim\,$$10^7$\,K) filling cosmic structures, which are detected in X-ray (e.g., \citealt{Anderson:2015} and refs.~within).   
Such plasma radiatively cools in the core ($< 0.1\,r_{100}$) through a {\it top-down condensation cascade} to dense warm gas ($\sim10^4$\,-\,$10^5$\,K; optical/IR\,-\,UV) and cold gas ($\lta 100$\,K; radio), subsequently raining toward the nuclear region ($< 10^{-3}\,r_{100}$). Via recurrent collisions, the condensed clouds are rapidly funneled toward the inner stable orbit ($r_{\rm ISCO}\approx3\,r_{\rm S}$). Such process is known as {\it chaotic cold accretion} (\citealt{Gaspari:2013_cca}; \S\ref{s:macro}). CCA has been independently probed by several observational works (e.g., \citealt{Werner:2014,David:2014,Voit:2015_nat,Tremblay:2016} and refs. therein).

General-relativistic, radiative magneto-hydrodynamic simulations (GR-RMHD) provide crucial constraints for the last stage of feeding (e.g., \citealt{Sadowski:2015,Sadowski:2016_thick}; \citealt{Sadowski:2017} -- SG17; \S\ref{s:micro}). Near the ISCO, the final drastic SMBH pull converts a fraction of the gravitational energy into mechanical output, ejecting most of the mass via {\it ultra-fast outflows} (UFOs). Such outflows re-heat the core, while entraining the ambient gas, in a self-regulated AGN feedback loop (Fig.~\ref{f:sketch}). In the paper companion to this work (SG17), we present and discuss in-depth the GR-RMHD simulations results, including the mechanical and radiative efficiencies.

In \S\ref{s:link}, we will quantitatively link the macro and micro properties of cosmic structures and SMBHs by using first principles, as mass and energy conservation, and by preserving minimal assumptions based on last-decade observations. The final equations provide the mass outflow rates and velocities at different scales (and for different phases).
In \S\ref{s:comp}, we compare the predictions with recent ionized, warm, and molecular outflow samples, and discuss the limitations.
In \S\ref{s:subgrid}, we discuss how to apply our model to other studies, as subgrid simulations, semi-analytic (SAM) studies, or observational interpretations. 
In \S\ref{s:disc}, we carefully discuss the limitations of the model and additional important features (as the duty cycle and the $M_\bullet - \sigma_\ast$ relation).
In \S\ref{s:conc}, we summarize the main points and conclude with future prospects.

\section{Large and small scales efficiencies}
\label{s:efficiencies}
\noindent
We highlight here 3 key regions which are central to our study (see Fig.~\ref{f:sketch} for a full diagram).

(i) The region closest to the SMBH {\it horizon}, $r \lta 5\,r_{\rm S}$ (a few ISCO radii), where gas is rushing toward the BH and there is no outflow. This region is fully resolved by the horizon scale GR-RMHD simulations. We denote properties in this region by a black dot, e.g., $\dot M_\bullet$. 

(ii) The ultra-fast outflow {\it launching} region, $r \lta 100\,r_{\rm S}$, within which the binding energy of the infalling gas is converted into mechanical outflow, not interacting yet with the ambient gas. We denote this by $\dot M_{\rm out}$.

(iii) The {\it macro} region, $r < r_{\rm core}\approx 10^9\,r_{\rm S}\approx 0.1\,{r_{\rm vir}}$ , within which the nuclear outflow is entrained (denoted by $\dot M_{\rm OUT}$), slowed down, and eventually thermalized (via bubbles, shocks, and turbulent mixing). 
The CCA rain develops in such core, with major collisions increasing within the kpc scale
(10\,-100\, Bondi radii\footnote{The Bondi radius, $r_{\rm B}=GM_\bullet/c^2_{\rm s,x}\simeq(7.5\,{\rm pc})\,M_{\rm \bullet, 9}\,T_{\rm x,7.4}^{-1}\approx10^5\,r_{\rm S}$, is not strictly relevant for CCA but provides a known reference intermediate (pc) scale between the macro and micro region.}). 

\subsection{Macro efficiency: chaotic cold accretion [CCA]} \label{s:macro}
\noindent
We introduce now the two key property of the feeding and feedback, i.e., the mechanical efficiency which has dimension of a power divided by the rest mass energy rate, $\varepsilon \equiv P/(\dot M c^2)$.

The best consistent way to solve the cooling flow problem appears to be
  mechanical AGN feedback self-regulated via CCA (\S\ref{s:intro}). 
  Solving the cooling flow problem means to avoid at the same time overcooling and overheating, preserving the inner structure of hot halos for $\sim10$\,Gyr,
  as tightly constrained over the last decade by {\it Chandra} and {XMM-\it Newton} data (e.g., \citealt{McNamara:2012,McDonald:2016}). Such hot halos are continuously perturbed by subsonic turbulence (e.g., \citealt{Khatri:2016}).
  In turbulent regions where the cooling time drops below the local dynamical time,
  nonlinear multiphase condensation develops (Fig.~\ref{f:sketch}, bottom).
  Such cold clouds and warm
  filaments collide in chaotic, inelastic way while raining on the SMBH (G15, G17; see also \citealt{Pizzolato:2010}), boosting the accretion rate with rapid intermittency.
  Massive sub-relativistic outflows are then triggered with kinetic power $P_{\rm OUT}$
  proportional to the large-scale inflow rate, preventing a run-away pure cooling flow (\S\ref{s:link}).

Due to self-regulation, the large-scale mechanical efficiency can be estimated by comparing the AGN energy output with the radiative energy losses, $P_{\rm OUT}\simeq L_{\rm x}$, yielding (\S\ref{s:in} for the derivation)
\begin{equation}\label{e:1}
\varepsilon_{\rm BH}\simeq10^{-3}\;T_{\rm x,7.4} \propto \frac{c^2_{\rm s,x}}{c^2},
\end{equation}
where $c_{\rm s,x}$ is the hot halo adiabatic sound speed and $c$ is the speed of light (the scaling shares analogy to a Mach number squared).
From less massive, lower-temperature, compact galaxies to more massive, hotter, and larger galaxy clusters ($T_{\rm x}\approx0.5-10$\;keV), the mechanical efficiency covers a range $\varepsilon_{\rm BH} \simeq 
2\times 10^{-4} - 4\times 10^{-3}$.
The macro efficiency is a function of hot halo temperature ($\propto T_{\rm x}$), thus total mass, decreasing for smaller halos since the cooling rate is a function $\propto L_{\rm x}/T_{\rm x}$ (as seen later in Eq.~\ref{e:6}). 
Smaller, less bound halos experience a stronger relative condensation due to the lower specific internal energy,
and necessitate of less sinked material -- with slightly more evacuation -- in order to avoid overheating.
Such quasi thermal equilibrium constraint on X-ray halos filling cosmic systems is key to set the macro efficiency. 

This picture has been corroborated by self-regulated AGN simulations of CCA and massive outflows tested in clusters, groups, and isolated galaxies (e.g., \citealt{Gaspari:2011a,Gaspari:2011b,Gaspari:2012b,Gaspari:2012a,Prasad:2015,Yang:2016}), which independently retrieve the same range of feedback efficiencies described above in varying systems.
The few available observational estimates, albeit limited by several extrapolations, are
also consistent with a mechanical efficiency of the order of $\varepsilon_{\rm BH}\sim10^{-3}$ (\citealt{Merloni:2008}).

\subsection{Horizon efficiency: GR-RMHD} \label{s:micro}
\noindent
Gas approaching the SMBH liberates its gravitational energy. A test particle falling straight on the BH would convert the liberated amount into kinetic energy of radial motion and, finally, take it with it below the horizon. From the point of view of the observer at infinity, no energy has been extracted.
Accretion flows act in a more complex way. The liberated gravitational energy goes mostly into kinetic motion. The turbulent nature of the flow induces this energy to dissipate and heat up the gas. At the same time outflows can be generated often via the magneto-centrifugal mechanism.
Only for idealized models, like advection dominated accretion flows (e.g., \citealt{Narayan:1995}), all the dissipated heat is advected with the flow onto the BH. In a more general case, energy is extracted from the system, and gas infalling from large radii and marginally bound, crosses the BH horizon with negative energy.

The amount of the extracted energy, i.e., the efficiency of a given accretion flow, depends solely on the energetics of the magnetized gas crossing the BH horizon. E.g., if on average gas with energy $e=-0.01\,\rho c^2$ falls into the BH, then the luminosity of such a system, as seen from infinity, is $L=0.01\,\dot M_{\bullet} c^2$. The properties of the accretion flow in the innermost region must be determined by numerical means, since the flow is highly nonlinear, strongly magnetized, and turbulent. 
In the companion paper, SG17, general relativistic radiative simulations of magnetized gas falling on the SMBH are carried out, testing over 5 orders of magnitude in accretion rates.
SG17 show that for a non-rotating BH and standard non-saturated configuration of the magnetic field, thick accretion flows (as expected in the maintenance mode of AGN feedback) have a fairly stable extraction of the rest mass energy accreted through the horizon,
\begin{equation} \label{e:2}
\varepsilon_{\bullet}\simeq0.03\pm0.01,
\end{equation}
Such mechanical efficiency will be the reference horizon efficiency for our model. 
We note that chaotic accretion (our macro scale model) will naturally lead to an average null spin configuration (e.g., \citealt{King:2006}).
An important result from SG17 is that such value is essentially independent of the ion-electron temperature ratio, i.e., the strength of the gas cooling does not affect the mechanical efficiency value at the micro scale. 

This energy outflow accelerates within the inner region ($\sim$\,$100\,r_{\rm S}$) and is ejected in a quasi-spherical way (Fig.~\ref{f:sketch}, top) in the form of an ultra-fast kinetic outflow of gas. The outflow is both thermally (equatorial) and magnetically driven (polar region; see also the simulations by \citealt{Moller:2015}).
At larger distances, the outflow interacts with the ambient medium, entraining gas via shocks and mixing instabilities, finally dissipating its energy within the core region, $r_{\rm c}\approx r_{500}/5 = (148\,{\rm kpc})\,T_{\rm x,7.4}^{1/2}$ (App.~\ref{app:L0}).
On top of this energy flux there might be a very thin, relativistic jet forming whenever the SMBH quickly spins and the magnetic field threads the horizon. Such a jet may be substantially energetic and could lead to larger efficiencies (e.g., \citealt{Tchekhovskoy:2011}). However, relativistic jets are in most cases very collimated and less likely to interact with the host. For such reason and for the null spin expected from chaotic accretion, 
we consider here only the wide, massive sub-relativistic outflows as dominant component of the kinetic feedback. 
Albeit not driving the total ram pressure, we note the jet and radio emission can still be correlated with the presence of massive AGN outflows, thus tracing some of the major AGN bubbles (\S\ref{s:disc}).

The emergence of ultra-fast outflows and the connection with large-scale warm absorbers has been corroborated by other analytic studies. \citet{Fukumura:2010,Fukumura:2014} show that magnetic torques acting on the inner rotating gas can efficiently drive an outflow through the magneto-centrifugal mechanism. The MHD wind is stratified, having slower velocity at progressively larger launching radii, akin to an entrained outflow.
In the radiatively efficient, Eddington regime, the spherical model by \citet{King:2014_WA} suggests that radiation pressure is able to drive UFOs; the expanding, swept-up shell is decelerated by the background medium, again corroborating the key role of the entrainment action in unifying AGN outflows over a large range of scales.

\section{Linking the macro and micro scales} \label{s:link}

\noindent
The two complementary simulations discussed above allow us to link the large-scale to small-scale properties of the feeding and feedback mechanism in a simple, coherent model. 
Fig.~\ref{f:sketch} illustrates the main features and characteristic scales of the model.

The large-scale outflow power can be modeled as
\begin{equation} \label{e:3}
P_{\rm OUT} = \varepsilon_{\rm BH} \dot M_{\rm cool} c^2,
\end{equation}
where $\dot M_{\rm cool}$
is the quenched cooling flow rate and $\varepsilon_{\rm BH}$ is the macro-scale mechanical efficiency (\S\ref{s:macro}).
The gaseous halo is losing internal energy via radiative emission (mainly via Bremsstrahlung), while the AGN feeds heating back, on average balancing the pure cooling flow.
Such halos perturbed by subsonic turbulence develop local multiphase condensation within the core,
as long as turbulent Taylor number ${\rm Ta_t}\equiv\sigma_v/v_{\rm rot} \lta 1$ (G15).
As cold clouds and filaments rain down, they experience recurrent chaotic, fractal collisions, which cancel angular momentum at progressively smaller radii, in particular as they collapse within $r<1$\,kpc.
The inflow rate can be thus considered independent of radius. 
In other words, during CCA rain, the cold gas condensed in the core is quickly funneled to the ISCO region with no long-term accumulation. G17 simulations showed that the CCA inflow rate is proportional to the effective viscosity of the cloud collisions, $\nu_{\rm c} \approx \lambda_{\rm c}\, \sigma_v$. The collisional mean free path $\lambda_{\rm c}$ and the ensemble velocity dispersion $\sigma_v$ are directly inherited from the large-scale turbulence (for a massive galaxy, $\lambda_{\rm c}\approx100$\,pc, $\sigma_v\approx150$\,km\,s$^{-1}$) -- a 3D chaotic process not tied to a radial dependence. 

The inner, tiny SMBH is the actual source of energy injection with power
\begin{equation} \label{e:4}
P_{\rm out} = \varepsilon_{\bullet} \dot M_\bullet c^2,
\end{equation}
where $\varepsilon_{\bullet}$ is the horizon efficiency (\S\ref{s:micro}) and $\dot M_\bullet$ is the inflow rate through the black hole horizon. The major difference between the macro and horizon efficiency implies that the sinked mass rate is the net inflow rate surviving the ultra-fast outflow generated near the ISCO scale, before falling into the unescaping BH horizon.\\

\subsection{Inflow properties} \label{s:in}
\noindent
The large-scale inflow rate is the {\it quenched} cooling flow rate. The maximal {\it pure} cooling flow (CF) rate 
can be calculated from the enthalpy variation of the hot gaseous halo via the first law of thermodynamics 
(e.g., \citealt{Gaspari:2015_xspec}) in isobaric mode, yielding
\begin{equation} \label{e:5}
L_{\rm x} = \frac{\gamma}{\gamma-1} \frac{k_{\rm b} T_{\rm x}}{\mu m_{\rm p}}\dot M_{\rm CF} = \frac{c^2_{\rm s,x}}{\gamma-1}\dot M_{\rm CF},  
\end{equation}
where $T_{\rm x}$ and $L_{\rm x}$ denote the core X-ray temperature and luminosity of the hot halo (App.~\ref{app:L0}), $\gamma=5/3$ is the adiabatic index, $\mu\simeq0.62$ is the average atomic weight for a fully ionized plasma with $\simeq25$\% He in mass, and $k_{\rm b}$ and $m_{\rm p}$ are the usual Boltzmann constant and proton mass, respectively. The last equality converts temperature into adiabatic sound speed, $c_{\rm s}= (\gamma{k_{\rm b} T_{\rm x}}/{\mu m_{\rm p}})^{1/2}\simeq1.5\times10^4\, T_{\rm x}^{1/2}$. From galaxies to clusters ($T_{\rm x}\approx0.5-10$\,keV), 
$c_{\rm s,x}=361-1615$\,km\,s$^{-1}$.

AGN feedback preserves the hot halos in quasi thermal equilibrium throughout the 10 Gyr evolution\footnote{\citet{McDonald:2017} show that cool cores are observed even up to $z\approx1.5$ with properties identical to local ones.}. The warm filaments drop out of the hot halo just below the soft X-ray regime (G17) as the cooling curve drastically increases due to line cooling. Thereby the actual mass flux arising out of the condensation process is linked to the suppressed soft X-ray luminosity. 
X-ray spectroscopical observations (e.g., \citealt{Peterson:2003,Kaastra:2004}) show that the soft X-ray emission is on average suppressed by 2 dex compared with the pure isobaric CF tied to the core $L_{\rm x}$ 
(cf.~\citealt{Gaspari:2015_xspec} for a review of observational works and analysis of the soft X-ray spectrum quenching). For such reasons, the effective quenched cooling rate is 
\begin{equation} \label{e:6}
\dot M_{\rm cool} \simeq 10^{-2}\,\dot M_{\rm CF} \simeq 6.7\times10^{-3} \frac{L_{\rm x}}{c^2_{\rm s,x}}.  
\end{equation}
We note such quenched, CCA rate is typically $100\times$ the Bondi rate (\citealt{Gaspari:2013_cca}), the latter being insufficient to properly boost the AGN heating (see also \citealt{Soker:2009,McNamara:2012}). 
Since hot halos are formed via the gravitational collapse of the cosmic structures, the temperature and luminosity are interchangeable via scaling relations (\citealt{Sun:2012}), 
such as $L_{\rm x}\simeq 6\times10^{43}\,(T_{\rm x}/{\rm 2.2\,keV})^3$ erg\,s$^{-1}$ (including the minor corrections due to the core radius instead of $R_{500}$; see App.~\ref{app:L0}).
We can thus rewrite Eq.~\ref{e:6} as
\begin{equation} \label{e:7}
\dot M_{\rm cool} \simeq (1.1\ {\rm \msun\,yr^{-1}})\,T_{\rm x, 7.4}^2 = (1.1\ {\rm \msun\,yr^{-1}})\,L_{\rm x, 43.8}^{2/3},
\end{equation} 
where the core $L_{\rm x}$ and $T_{\rm x}$ are in unit of $6\times10^{43}$\,erg\,s$^{-1}$ and $2.6\times10^7$\,K (2.2\,keV), respectively.
From compact galaxies to massive clusters ($T_{\rm x}\approx0.5-10$\,keV), the inflow rate covers 
$\dot M_{\rm cool}\simeq0.06-23\ {\rm \msun\,yr^{-1}}$.
Interestingly, all the below scalings can be also expressed in terms of total mass or virial radius, e.g., $\dot M_{\rm cool} \propto r_{\rm vir}$ (App.~\ref{app:L0}). It is important to note that if the core cooling time is $t_{\rm cool}\gta t_{\rm H}/2$, then the system is in a non-cool-core condition and no condensation rain, feeding, and feedback shall be applied (regardless of scaling relations), until the core cools down, igniting the self-regulated loop (see \S\ref{s:disc}).

The energy conservation requirement, 
\begin{equation}\label{e:8}
P_{\rm out}=P_{\rm OUT},
\end{equation} 
implies that the horizon inflow rate is related to the cooling rate as follows:
\begin{equation} \label{e:9}
\dot M_\bullet = \frac{\varepsilon_{\rm BH}}{\varepsilon_{\bullet}} \,\dot M_{\rm cool},
\end{equation}where 
the horizon mechanical efficiency is directly provided by the GR-RMHD simulations (\S\ref{s:micro}), $\varepsilon_{\bullet}=0.03$.
From the results and observations discussed in \S\ref{s:macro}, hot halos must avoid at the same time overheating and overcooling, i.e., the energy lost via radiative emission in the core must be replaced by the SMBH feedback power,
\begin{equation} \label{e:10}
P_{\rm OUT} \simeq L_{\rm x}.
\end{equation}
Thereby $\varepsilon_{\rm BH} = L_{\rm x}/(\dot M_{\rm cool}c^2)$ and by
using Eq.~\ref{e:6}, the macro efficiency reduces to
\begin{equation} \label{e:11}
\varepsilon_{\rm BH}=\frac{10^2}{\gamma-1}\frac{c^2_{\rm s}}{c^2}\simeq10^{-3}\;T_{\rm x,7.4}=10^{-3}\;L_{\rm x,43.8}^{1/3}
\end{equation}
Notice that the efficiency only mildly varies with the main variable, the X-ray luminosity.
We can now use both efficiencies to retrieve the horizon inflow rate relative to the macro value via Eq.~\ref{e:9} as
\begin{equation} \label{e:12}
\dot M_\bullet \simeq(0.03\,\dot M_{\rm cool})\;T_{\rm x,7.4} = (0.03\,\dot M_{\rm cool})\;L_{\rm x,43.8}^{1/3},
\end{equation}
i.e., only a few percent of the quenched cooling flow rate is actually sinked through the SMBH horizon. 
Substituting $\dot M_{\rm cool}$ in Eq.~\ref{e:12}, we consistently retrieve 
the accretion rate directly proportional to the X-ray luminosity, 
\begin{align} \label{e:13}
\dot M_\bullet &= \frac{L_{\rm x}}{\varepsilon_{\bullet} c^2}\simeq (0.04\ {\rm \msun\,yr^{-1}})\ L_{\rm x, 43.8} \\
&= (0.04\ {\rm \msun\,yr^{-1}})\ T_{\rm x, 7.4}^3. \nonumber
\end{align}
For SMBHs in the local universe, such accretion rates are typically sub-Eddington, as expected for the maintenance, mechanically dominated mode of AGN feedback.
As shown by \citet{Russell:2013} and corroborated by SG17, the radiative efficiency and thus power due to radiation is several dex lower than the mechanical input, and it can be neglected in terms of driver of the dynamics (albeit radiation is clearly relevant to detect and trace AGN; \S\ref{s:disc}). Eq.~\ref{e:12}-\ref{e:13} imply that SMBHs in lower mass halos have typically a lower {\it absolute} accretion rate. Moreover, a {\it relatively} smaller fraction of gas reaches the horizon as AGN feedback is more effective in halos with lower binding energy, which are tied to both lower $M_{\rm 500}$ and lower black holes masses (\S\ref{s:disc}).

\subsection{Outflow properties} \label{s:out}
\noindent
Having assessed the inflow properties, we are now in a position to retrieve the structure of the outflows, again via minimal first principles. The power in terms of characteristic mass outflow rates\footnote{The term due to $v\dot v$ is subdominant and can be neglected.}
and velocities at the launching and macro scale is
\begin{equation} \label{e:14}
P_{\rm out} = \frac{1}{2}\dot M_{\rm out}\, v^2_{\rm out},
\end{equation}
\begin{equation} \label{e:15}
P_{\rm OUT} = \frac{1}{2}\dot M_{\rm OUT}\, v^2_{\rm OUT},
\end{equation}
respectively.
As shown in Eq.~\ref{e:12}, only a few percent of the total inflow is actually sinked through the SMBH horizon; most of the mass is returned as ultra-fast outflows launched within $\sim$\,$100\,r_{\rm S}$, such as
\begin{align} \label{e:16}
\dot M_{\rm out} &= \dot M_{\rm cool}-\dot M_\bullet = \left(1-\frac{\varepsilon_{\rm BH}}{\varepsilon_{\bullet}}\right)\,\dot M_{\rm cool}\approx\dot M_{\rm cool},
\end{align}
which leads to the inner outflow velocity via Eq.~\ref{e:14}
\begin{align} \label{e:17}
&v_{\rm out} = \sqrt{\frac{2\,\varepsilon_{\bullet}\dot M_\bullet c^2}{\dot M_{\rm out}}} = \sqrt{\frac{2\,\varepsilon_{\rm BH}}{1-\varepsilon_{\rm BH}/\varepsilon_{\bullet}}}\;c\simeq \sqrt{2\,\varepsilon_{\rm BH}}\;c\\
&\simeq (1.4\times10^4\; {\rm km\,s^{-1}})\; T_{\rm x,7.4}^{1/2} = (1.4\times10^4\; {\rm km\,s^{-1}})\; L_{\rm x,43.8}^{1/6}. \label{e:18}
\end{align}
We note $v_{\rm out}$ in Eq.~\ref{e:17} can be tied to a momentum $p_{\rm out} = \dot M_{\rm out}\,v_{\rm out}$, which satisfies $(1/2)\dot M_{\rm out} v_{\rm out}^2=p_{\rm out}^2/(2\dot M_{\rm out})$.

Together with the above outflow rates, these are the typical velocities of ultra-fast outflows (UFOs) observed as blue-shifted absorption lines tracing the inner launching region near the SMBH gravitational radius (\citealt{Tombesi:2012,Tombesi:2013,Fukumura:2015};  more discussions and comparisons in \S\ref{s:comp}). We note the outflow velocity is only weakly dependent on the halo temperature/luminosity, varying at best by a factor of 2.5. We thus expect $10^4$\,km\,s$^{-1}$ to be a fairly general attribute\footnote{This is also similar to the characteristic nuclear (100\,-\,200\,$r_{\rm S}$) escape velocity, i.e., as the driven outflow overcomes gravity.}
of inner launching outflows (\citealt{Crenshaw:2003,Tombesi:2016} for reviews).

As the inner ultra-fast outflow propagates outward ($r\gg100\,r_{\rm S}$), it will entrain the {\it background} gas (embedding the low volume-filling CCA rain\footnote{Through the feedback cycle, the underlying halo gently expands during entrainment, and contracts after dissipation, restoring quasi hydrostatic equilibrium (HSE). X-ray observations indeed show that density profiles in cool-core systems vary only by a small amount, even after strong outbursts (e.g., \citealt{McNamara:2016}).}) along its way such as 
\begin{equation} \label{e:19}
\dot M_{\rm OUT} = \eta\,\dot M_{\rm out},
\end{equation}
where $\eta > 1$ is the entrainment factor.
We note at kpc scale the mechanical outflow has not yet thermalized, conserving most of the kinetic energy, as we see the formation of X-ray cavities and hot spots at larger distances.
At a given radius, the entrained mass outflow rate can be retrieved via the mass flux equation 
\begin{align}
\dot M_{\rm OUT} &= \Omega\, r^2\,\rho(r)\, v_{\rm OUT}(r) = \Omega\, r^{2-\alpha}\rho_0 r_0^{\alpha}\;\,v_{\rm OUT}(r) \nonumber\\
& \simeq \Omega \rho_0 r_0\;r\,v_{\rm OUT}, \label{e:20}
\end{align}
where the inner gas density profile is typically a power-law $\rho = \rho_0 (r/r_0)^{-\alpha}$
and $\Omega\le 4\pi$ is the covering angle of the bipolar outflow. 
As shown in G17 and observational refs. within, the typical nuclear profiles for all the phases follow a slope $\alpha\simeq1$ (with $\approx$\,0.25 scatter), hence the last step in Eq.~\ref{e:20}.
By using Eq.~\ref{e:15} and \ref{e:19}, the entrained outflow velocity can be written as
\begin{equation} \label{e:21}
v_{\rm OUT} = \sqrt{\frac{2 P_{\rm OUT}}{\dot M_{\rm OUT}}}=\eta^{-1/2}\,v_{\rm out},
\end{equation}
which inserted in Eq.~\ref{e:20} yields an entrainment factor
\begin{equation} \label{e:22}
\eta = \left(\Omega\rho_0 r_0\:r\,\frac{v_{\rm out}}{\dot M_{\rm out}}\right)^{2/3} \propto \frac{r^{2/3}}{T_{\rm x}}. 
\end{equation}
This implies that, while the macro velocities at a given radius are unchanged over different systems ($v_{\rm OUT}\propto T_{\rm x}^{1/2}/T_{\rm x}^{1/2}$), and are thus more robust probes, the macro outflow rate linearly increases for more massive systems $\dot M_{\rm OUT}\approx \eta\,\dot M_{\rm cool}\propto T_{\rm x}$.
Note the mass outflow rate has much stronger relative variations than velocities ($\propto \eta^{-1/2}$), corroborating Eq.~\ref{e:14}-\ref{e:15}.

Depending on the current thermodynamical background state of the system, the outflows can entrain different phases, including the hot plasma, the warm neutral/ionized gas, and the molecular gas.
We use the results of the CCA simulations (G17) to retrieve the multiphase environment and profiles of the 3 phases, taking as reference macro scale $r_0=1$\,kpc.
A typical plasma density
$\rho_{\rm 0, hot}\simeq10^{-25}$\,g\,cm$^{-3}$ at 1 kpc leads to an entrainment factor ($\Omega \simeq 4\pi$)
\begin{equation} \label{e:23}
\eta_{\rm hot} \simeq 40\ T_{\rm x,7.4}^{-1}\,r_{\rm 1\,kpc}^{2/3}.
\end{equation}
This implies median entrained mass outflow rates and velocities 
of 10s\,$\msun$\,yr$^{-1}$ and a few $10^3$\,km\,s$^{-1}$, which are typical properties of observed macro ionized outflows (e.g., \citealt{Nesvadba:2010,Tombesi:2013}). 
If the halo is mainly filled with cooler gas, such as at high redshift, 
the entrainment can also proceed mainly via the warm ($\rho_{\rm 0, warm}\simeq10^{-24}$\,g\,cm$^{-3}$) and cold phase ($\rho_{\rm 0, cold}\simeq10^{-23}$\,g\,cm$^{-3}$)\footnote{Here we assume that the characteristic phase densities retrieved in G17 apply over the whole inner region as a background; this is more typical at high redshift, as cold flows can penetrate deep within the growing proto-galaxy.}, 
 thus leading to more entrained outflows with
\begin{equation} \label{e:24}
\eta_{\rm warm} \simeq 183\ T_{\rm x,7.4}^{-1}\,r_{\rm 1\,kpc}^{2/3},
\end{equation}
\begin{equation} \label{e:25}
\eta_{\rm cold} \simeq 850\ T_{\rm x,7.4}^{-1}\,r_{\rm 1\,kpc}^{2/3}.
\end{equation}
Mass outflow rates with $10^2$ and several $10^2$\,$\msun$\,yr$^{-1}$ tied to velocities $1000$ and 500\,km\,s$^{-1}$ at the kpc scale are characteristic properties found throughout observations of 
neutral (e.g., \citealt{Morganti:2005,Morganti:2007,Teng:2013,Morganti:2015_rev}) 
and molecular AGN outflows (\citealt{Sturm:2011,Cicone:2014,Russell:2014,Combes:2015,Feruglio:2015,Morganti:2015_ALMA,Tombesi:2015}), respectively (more detailed comparisons in \S\ref{s:comp}).

At large radii, the outflow is halted by the external pressure, inflating a bubble and thermalizing its kinetic energy mainly via turbulent mixing (e.g., \citealt{Gaspari:2012b,Soker:2016,Yang:2016}). Such radius crudely corresponds to the region where the outflow ram pressure becomes equal to the hot halo pressure. Since outflow ram pressure is equal for all the phases, we can estimate the thermalization radius as $v_{\rm OUT,\rm hot}^2\sim c_{\rm s}^2/\gamma$, yielding via Eq.~\ref{e:21}-\ref{e:23}
\begin{equation} \label{e:26}
r_{\rm th}\sim ({\rm 55\,kpc})\; T_{\rm x,7.4}^{3/2} = ({\rm 55\,kpc})\; L_{\rm x,43.8}^{1/2}. 
\end{equation}
Above such thermalization radius, any model should simply inject thermal energy rate balancing the core $L_{\rm x}$.
Below such radius (as resolved by most of the current MHD and cosmological simulations), any model should inject massive outflows with the above relations.
Such radius roughly approaches the core radius, which is where the feedback loop is active.

In principle the momentum equation, $\dot M_{\rm OUT}\,v_{\rm OUT}=\dot M_{\rm out}\,v_{\rm out}$, might be adopted instead of Eq.~\ref{e:14}\,-\,\ref{e:15}, if the outflow would immediately radiate away most of its energy. However,
besides losses being likely subdominant (see \citealt{Faucher:2012}), the deceleration would result to be dramatic,
$v_{\rm OUT}=v_{\rm out}/\eta$ (with $\eta\propto r^{1/2}\,T_{\rm x}^{-3/4}$ reduced by a few), which would make the outflow aborted at the macro scale, inconsistently with data. Adopting the same procedure as above, the hot, warm, and molecular outflow would merely preserve 870, 280, and 90 km\,s$^{-1}$ at 1 kpc scale, respectively. A related crucial point to reject purely momentum-driven outflows is that self-regulation would be broken, since the macro feedback energy could not balance the core $L_{\rm x}$, leading to a global massive pure CF.

\section{Comparison with observations} \label{s:comp}
\noindent
The proposed CCA GR-RMHD unification predicts nuclear ultra-fast outflows of the order of $10^4$\,km\,s$^{-1}$ and a progressively slower propagation of the outflow at larger radii, which are consistent with recent AGN data. 

\begin{figure}
\centering
\includegraphics[width=\columnwidth]{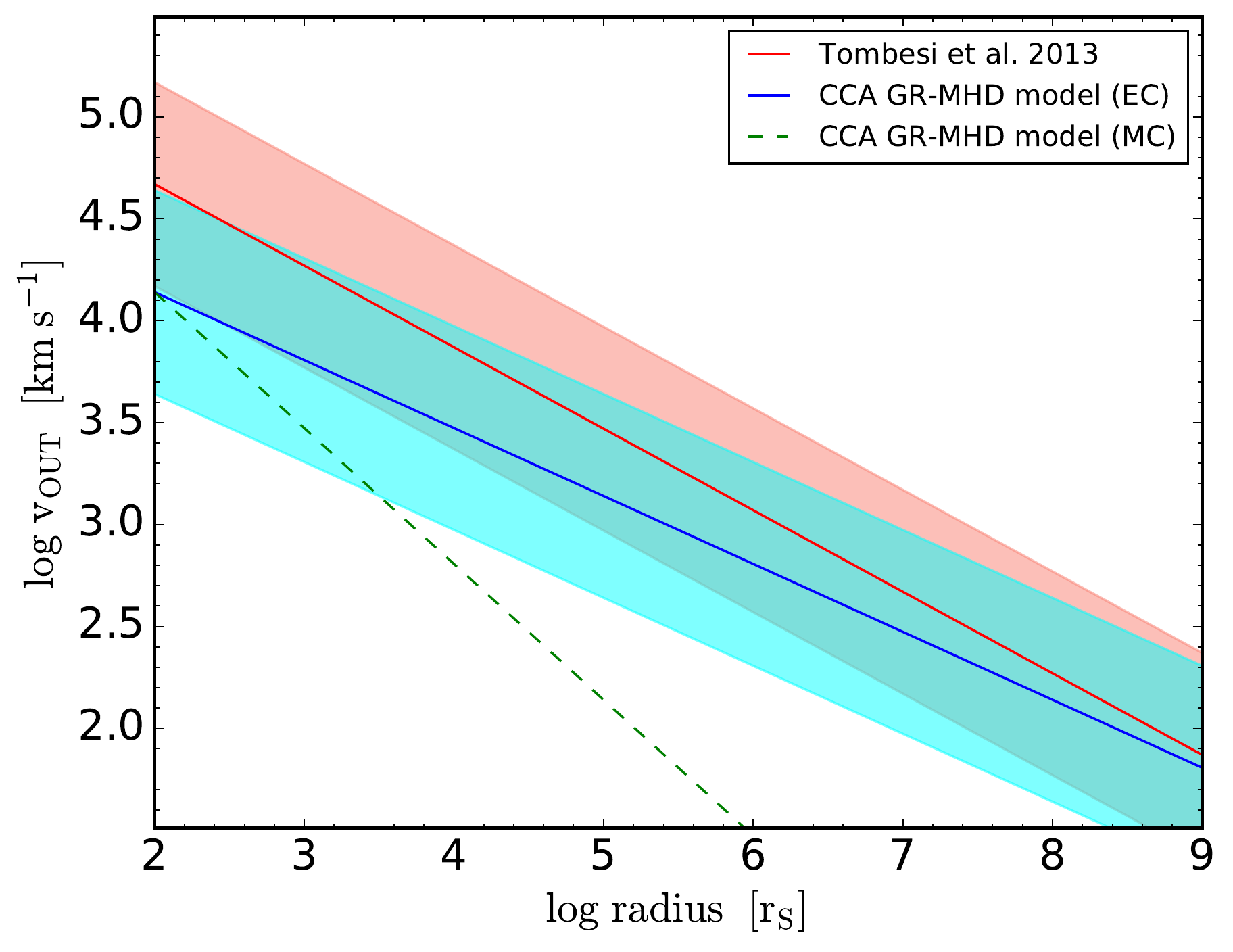}
\caption{Outflow velocity as a function of radial distance (normalized to the Schwarzschild radius) for the unified X-ray UFO plus warm absorber data (red; \citealt{Tombesi:2013}) and the prediction of our energy-conserving CCA GR-RMHD model (blue; \S\ref{s:link}). The dashed green line shows the (inconsistent) purely momentum-driven outflow. The region within $\sim$\,$100\,r_{\rm S}$ is the UFO generation region, where most of the inflow mass is ejected. At larger radii, the UFO entrains progressively more mass, slowing down. The adopted profile slope of the warm gas background is $\alpha=1$.
The proposed model, based on linking the horizon/GR-RMHD and macro/CCA efficiencies, well reproduces the data within scatter.} 
\label{f:comp}
\end{figure}

In a sample of 35 AGN, \citet{Tombesi:2013} unify the velocities of UFOs and the slower warm absorbers as a function of radial distance (see also \citealt{Tombesi:2014} for analogous radio galaxy sample). Velocity is the most robust indicator (e.g., compared to mass outflow rates) since directly observed through blue-shifted absorption lines in AGN X-ray spectra. Fig.~\ref{f:comp} shows the comparison of our model prediction (blue; \S\ref{s:link}) and the fit to the unified X-ray data. The bands denote 0.5\,dex scatter, which is the typical model variation (mainly due to inner density and bipolar angle) 
and the range in the observed data points.
The prediction of the CCA GR-RMHD model well reproduces the observed values. If the outflow would be purely driven by momentum (green line) and not energy, it would be aborted within the Bondi radius, remaining clearly below data. In other words, entrainment must occur in a gentle way, such as $v_{\rm OUT}\propto \eta^{-1/2}\propto r^{-1/3}$. 
In the nuclear region, the outflow tends to be slightly lower than the data, albeit within typical uncertainties. The slope of the data, -0.40, is slightly steeper than the -0.33 model. The two matches exactly if the density profile has slightly shallower $\alpha=0.8$ (instead of 1); we did not attempt to fine-tune it, since within uncertainties of the simulated radial profiles and not granting further insight.

The mass outflow rates have very large observational uncertainties (due to the unknown geometry and projection effects) and theoretical scatter (due to the $T_{\rm x}$ dependance, unlike the macro velocity).
In the above sample, UFOs typically show $\dot M_{\rm out}\approx0.3\,\msun\,{\rm yr^{-1}}$, while the warm absorbers have 1.5\,-\,2 dex larger magnitude, which can be explained via the entrainment action ($T_{\rm x}\approx0.6$\,keV).
We are here not attempting to fit values of single objects; nevertheless, several X-ray studies detect nuclear $10^4$\,km\,s$^{-1}$ UFOs and ionized outflows with $10^{3}$\,km\,s$^{-1}$ at intermediate scale down to several 100 km\,s$^{-1}$ at large radii\footnote{In low-mass galaxies the thermalization radius is $<10$\,kpc, thus the outflow can rapidly decline in velocity (and mass rate).} (see the review by \citealt{Tombesi:2016} and refs.~within).
Follow-up observational investigations are required to better unify the radial properties of ionized outflows over a large homogeneous sample, in particular adding more low-luminosity AGN and central galaxies.

Depending on the dominant nuclear phase, the AGN ejecta can also develop into a neutral and molecular outflow. This is more common in QSOs and ULIRGs with abundant cold/warm mass with large volume filling in the core.
\citet{Morganti:2005, Morganti:2007} have shown the incidence of HI outflows in several AGN, particularly radio-loud sources, via (21-cm) radio telescopes as WRST.
The location of the HI outflows is 0.5\,-\,1.5\,kpc with average velocities
1000\,km\,s$^{-1}$. 
 \citet{Teng:2013} present a sample of 27 kpc-scale HI outflows detected with GBT: the  average sample velocity is $v_{\rm OUT, neu}\simeq 885$ km\,s$^{-1}$. In both samples, the mass outflow rates are uncertain (due to the dynamical time estimate), of the order of 100\,$\msun$\,yr$^{-1}$.
The above values are consistent with our median prediction of neutral outflows (Eq.~\ref{e:21}-\ref{e:24}) with a typical $v_{\rm OUT, neu}\simeq 1035$\,km\,s$^{-1}$ and $\dot M_{\rm OUT, neu}\simeq 92$\,$\msun$\,yr$^{-1}$ at kpc scale ($T_{\rm x}\approx1$\,keV).

In the last several years and with the advent of high-resolution radio interferometers, neutral outflows have been complemented with samples of massive molecular AGN outflows. 
\citet{Cicone:2014} present a sample of 19 molecular AGN outflows detected with IRAM (by using CO[1-0] emission closely tracing H$_2$ gas) at the kpc scale. Averaging the peak velocity and mass outflow rates over the sample yields a velocity $v_{\rm OUT, mol}\simeq 573$ km\,s$^{-1}$ 
and mass rate $\dot M_{\rm OUT}\simeq 428\ {\rm \msun\;yr^{-1}}$ with factor of 2 uncertainty.
The sample of 6 molecular outflows in \citet{Sturm:2011} show very similar mean properties. 
From Eq.~\ref{e:25}, the average molecular velocity and mass outflow rate at kpc scale is expected to be 480\,km\,s$^{-1}$ and $\dot M_{\rm OUT}\simeq 425\ {\rm \msun\;yr^{-1}}$ ($T_{\rm x}\approx$\,1\,keV), in agreement with the data.
Other works focus on single objects, finding very similar properties at kpc scale as predicted by our model; e.g., Phoenix/A1664 BCG cores display $v_{\rm OUT, mol}\simeq 550/590$ km\,s$^{-1}$ and crude outflow rates $>250\ {\rm \msun\;yr^{-1}}$ (\citealt{Russell:2014,Russell:2016_phoenix}). 
A well studied multiphase outflow in both the hot and cold phase is Mrk 231 (\citealt{Feruglio:2010, Feruglio:2015}).
IRAM data indicates a kpc-scale molecular outflow with $v_{\rm OUT, mol}\simeq 750$ km\,s$^{-1}$ and $\dot M_{\rm OUT}\simeq 700\ {\rm \msun\;yr^{-1}}$ (\citealt{Feruglio:2010}); in the same system, {\it Chandra} and NuSTAR show the presence of a nuclear UFO with $v_{\rm out, hot}\simeq 2\times10^{4}$ km\,s$^{-1}$ and $\dot M_{\rm OUT}\simeq 1\ {\rm \msun\;yr^{-1}}$. Both values are in excellent agreement with our entrainment multiphase model. Notably, the same authors remark that energy is conserved during the entrainment process, $P_{\rm out}\approx P_{\rm OUT}$, consistently with our Eq.~\ref{e:8}.
\citet{Tombesi:2015} present another similar multiphase outflow in IRAS F11119+3257.
As above, the mass outflow rates bear large uncertainties and a large sample linking the small and large radii (as done for UFOs) is currently missing; we encourage observational proposals in such unification direction. We are living a new era for multiphase AGN outflows, as the field is rapidly growing via new high-resolution ALMA cycles able to probe $\sim500$\,km\,s$^{-1}$ CO outflows (as shown by \citealt{Morganti:2015_ALMA}).

\section{Subgrid/SAM model for AGN feedback} \label{s:subgrid}
\noindent
Below we describe how to incorporate our model into large-scale simulations of structure formation. Let us denote the typical resolution of a given simulation by $\Delta r$ (nowadays $\sim 1{\rm\, kpc} \gg r_{\rm S}$ in a typical zoom-in run). Assuming the resolution is enough to resolve the thermalization region ($\Delta r <  r_{\rm th}$), we propose the following.

(i) The SMBH growth can be tracked via Eq.~\ref{e:12} or \ref{e:13},
\be \nonumber
\dot M_{\rm \bullet}\simeq (0.03\,\dot M_{\rm cool})\,T_{\rm x,7.4} = (0.04\ {\rm \msun\,yr^{-1}})L_{\rm x, 43.8},
\ee
i.e., only a few percent of the macro cold inflow rate is actually deposited into the SMBH (with coarse resolution it may be easier to estimate the cold inflow from the core $L_{\rm x}$, with the condition that the current central cooling time is shorter than $t_{\rm H}/2$; see Eq.~\ref{e:7}).

(ii) The AGN mechanical feedback is injected on the scales defined by a few $\Delta r$ with velocity given by Eq.~\ref{e:21},
\be \nonumber
v_{\rm OUT}=\eta^{-1/2}\,v_{\rm out},
\ee
where $\eta(r\approx\Delta r)$ is the entrainment factor at the resolved radial distance (Eq.~\ref{e:22}) and $v_{\rm out}$ is the nuclear velocity of the outflow set by Eq.~\ref{e:17}.

(iii) The rate at which such outflow carries mass results from the entrainment mechanism given by Eq.~\ref{e:19},
\be \nonumber
\dot M_{\rm OUT}=\eta\,\dot M_{\rm out} \simeq \eta\, \dot M_{\rm cool},
\ee
where $\dot M_{\rm cool}$ reflects the magnitude of the quenched cooling flow, which should self-consistently arise from the AGN feedback loop as a central cold inflow.
The outflow can be injected as a mass flux through the boundary (e.g., sink the inflow rate and inject it back boosted by a factor $\eta$).
If resolution does not permit to resolve the CCA inflow,
it is better to not sink the gas and kick the gas mass per timestep over the most inner number of cells/particles (reaching $\dot M_{\rm OUT}$) directly in the domain (checking for stability). Such inner active mass per timestep is naturally a fair representation of the entrained mass outflow rate
(as tested in \citealt{Gaspari:2011b,Gaspari:2012b}). 
A remark is that for very coarse resolutions $\Delta r > r_{\rm th}$ (Eq.~\ref{e:26}), injecting massive outflows loses physical meaning, and 
the average radiative energy losses should be simply balanced via thermal energy injection, since the outflows are expected to be thermalized.

Such prescription is perfectly suited to be used also in semi-analytic models (SAM), e.g., of galaxy and cluster evolution, as well as in the interpretation of observational data (limited by the instrumental -- instead of numerical -- resolution).
Furthermore, the injected properties, in particular the efficiency, are known a priori, regardless of numerics, implying that the fine-tuning loop plaguing current cosmological runs can be avoided (typically fitting one mass range, but overheating or overcooling the opposite regime due to keeping a constant macro efficiency). In other words, there is no main free parameter involved, except for the scatter intrinsic in observations. A sanity check is to retrieve the observed X-ray properties, e.g., X-ray luminosity and temperature profiles of the group or cluster. If not, the implementation of AGN feedback is numerically flawed and shall be modified accordingly, not retuning the parameters, but changing the injection implementation and carefully assessing which hydrodynamic solver and discretization to use. In other words, retuning some parameters to counteract the numerical flaws must be avoided, and can be avoided with the above a-priori prescription, thus preserving predictability.

\section{Discussion} \label{s:disc}
\noindent
We now discuss some details of the proposed model, together with the limitations and possible improvements.

The approach of this work differs from typical analytic modeling considering a perfect steady state solution (e.g., \citealt{Bondi:1952}) in which inflow and outflow coexist at exactly the same time (setting $\partial/\partial t = 0$ in the hydro equations).
As indicated by X-ray observations and simulations (\S\ref{s:intro}), the detailed self-regulated AGN feedback loop is time varying.
We have instead considered a nearly stationary case over a feedback cycle, which is typically of the order of the central cooling time $t_{\rm cool}\simeq k_{\rm b} T/(n \Lambda)$, where $\Lambda$ is the plasma cooling function (\citealt{Sutherland:1993}); from isolated galaxies to massive clusters the typical central cooling time of the hot gas varies from tens to several 100 Myr (\citealt{Gaspari:2014_scalings}). Within one cycle the process is time varying, with energy and mass changing form and phase.
Specifically, the inflow acts first via the self-similar CCA rain, then the SMBH reacts to the feeding via nuclear ultra-fast outflows (Fig.~\ref{f:sketch}). 
The propagating UFOs entrain the diffuse phase and thermalize in the core, such that $P_{\rm OUT}\simeq L_{\rm x}$, as shown by X-ray data (e.g., \citealt{Main:2017}).
The background halo is recurrently contracting and expanding in a gentle manner, and is never evacuated; in other words the core oscillates near HSE. 
Over the whole core region and one loop time the mass and energy are conserved (the small mass loss onto the BH is replenished from the virial hot halo).
Note that if central $t_{\rm cool}\gta t_{\rm H}/2$, the system is in a non-cool-core condition and the feeding/feedback is not currently active.
A key observational evidence for a variable feeding mechanism, is the ubiquitous variability of AGN light curves. As discussed in G17 (Sec.~5.1) and \citet{King:2015_flicker}, chaotic accretion drives a `flicker' noise with major accretion events having Myr duration.

Needless to say a full, time-dependent treatment of the feeding and feedback process requires 3D (GR)MHD simulations covering the whole dynamical spatial and temporal range. However, until we will be able to break such computational barrier, we can rely on key properties of the inflows and outflows set by the multiwavelength constraints, which must be satisfied even in the advanced numerical runs.
We remark X-ray data show that the feedback must be gentle and kinetically driven (with large-scale thermalization up to 100s kpc for massive clusters).
Notice that the details of the energy conserving outflow are in our macro model not relevant. On the other hand, the momentum flux boost of the swept-up material due to the hot shocked gas and entrainment via hydro instabilities (e.g., Kelvin-Helmholtz and Rayleigh-Taylor) requires numerical simulations to be robustly understood. In addition to direct uplift, an interesting possibility to form molecular outflows is the in-situ condensation of the massive galaxy-scale hot wind via thermal instability -- as discussed by \citet{Zubovas:2014} -- which may further promote the subsequent precipitation phase.

In this work, we decided to aim for minimal assumptions and rely on first principles as much as possible.
Further sophistications to the model are possible and can be easily incorporated to fit more specific objects, at the expense of an increased number of parameters.
For instance, the inner background density profile can be modified with a more complex functional form than a single power law and/or assigning different volume filling profiles to the warm/cold phases.
The configuration of the inner outflows can be modified by reducing $\Omega$, in order to accommodate for a thinner bipolar setup.
We note, in one loop, the cold inflow can occur along one direction, while the entrained outflow may occur in the perpendicular direction, further corroborating the separation of the large-scale CCA inflow and outflow mass rate, instead of a perfectly radial steady-state solution. A time delay in the loop can be introduced by tracking turbulent Taylor number: if $\rm Ta_t > 1$, then a rotating structure (disc, ring, torus) can momentarily reduce accretion.
We did not aim to fit one particular system or AGN outflow in this study, discussing only mean values. As noted in \S\ref{s:comp}, considering the scatter in cooling system properties, the outflow variations are $\sim$\,0.5~dex over a large sample. Fitting and interpreting single object data can be easily refined, e.g., by analyzing the core and nuclear X-ray spectrum both in terms of cooling rate (soft X-ray) and outflow line absorption features (hard X-ray). 

Consistently with the observational results by \citealt{Russell:2013} (Fig.~12), the GR-RMHD simulations (SG17) show that for accretion rates below $10^{-2}$ the Eddington rate, the nuclear SMBH power is dominated by kinetic energy over the SMBH radiative output, $P_{\rm out}\gg L_{\rm AGN}$. The mechanical, sub-Eddington mode is the long-term maintenance mode of AGN feedback (\citealt{McNamara:2012} for a review) preserving hot halos and cool-core systems in quasi thermal equilibrium at least for 9\,-10\,Gyr (\citealt{McDonald:2014,McDonald:2016,McDonald:2017}).
At high redshift ($z>2$), the Eddington rate can be approached triggering a brief `quasar' phase (seeding part of the SMBH mass). The wind may be thus radiatively driven, although its coupling with the gas is matter of ongoing debate. Moreover, there is no physical reason to think that the mechanical power from AGN is erased in this regime, as corroborated by our GR-RMHD run covering the quasar transition (see SG17). Even in such short-lived radiative regime, the outflow is still expected to be energy conserving\footnote{As cooling acts on electrons, this slows down inverse Compton process; free-free cooling is secondary.} (\citealt{Faucher:2012}) although it may be more appropriate to use a slightly larger $\varepsilon_{\bullet}\simeq0.057$ (\citealt{Novikov:1973,Merloni:2008}) and rescale Eq.~\ref{e:9}. As long as $ \varepsilon_\bullet \gg \varepsilon_{\rm BH}$, the outflow properties are however not significantly altered. 
A few quasar blasts may evacuate the system, but these anomalously powerful outbursts -- which are much easier to detect -- must be outliers (increasing the high-redshift population scatter) otherwise the majority of systems would later remain non-cool-core, which is not observed (e.g., \citealt{Gaspari:2014_scalings}).
Overall, regardless of the details of the driving mechanism (e.g., magnetic versus radiative), if self-regulation is on average preserved, the proposed model applies in similar way throughout cosmic time. 

In the current interpretation, the micro and macro mechanical driver is a sub-relativistic outflow. Given the BH null spin expectation from chaotic accretion (\citealt{King:2006}) and the high piercing collimation, a radio jet is expected to be subdominant, albeit it can coexist and trace the large-scale features, as bubbles. Observationally, radio synchrotron (electron) power is less than a percent of the cavity internal power (\citealt{McNamara:2012}), so only relativistic ions are left to inflate a bubble; however, this would produce strong Gamma emission in all systems, which {\it Fermi} does not typically observe. Moreover, several AGN bubbles are ghost cavities devoid of radio emission.
Having said that, our model is general and the radio jet interpretation can be trivially implemented, e.g., by replacing the related micro efficiency and opening angle. 

A current observational limitation which is worth discussing is the low-mass end regime.
While hot, X-ray halos are well detected above stellar masses $M_\ast \gta10^{10.8}\,\msun$, in particular massive galaxies, galaxy groups and clusters, the precise level of the X-ray luminosity due to the diffuse component in the opposite regime ($T_{\rm x}\lta 0.3$\,keV) is still uncertain due to the contamination of X-ray binaries (e.g., \citealt{Anderson:2015}). 
The X-ray luminosity in such regime may be lower than our adopted scaling, and the relative cooling rate (Eq.~\ref{e:7}) should be properly rescaled if necessary. 
While the outflow velocities are overall unaffected (Eq.~\ref{e:18}-\ref{e:21}), the mass outflow rate may be lower than the expected value.
Conversely, while more massive systems have better constrained core X-ray luminosity, the stronger and harder diffuse emission substantially hinders the nuclear X-ray spectral features, making UFO detection challenging.
If $T_{\rm x}$ is not available (e.g., for proto-galaxies), we suggest to use a core temperature in lower energy bands, as condensation occurs throughout the warm and cold phase regime.
Finally, supernova feedback due to star formation (e.g., with rate a few precent of the galaxy cooling rate) can also become energetically important in low mass galaxies and shall be investigated in the future.

While here we have investigated the instantaneous properties as the SMBH accretion rates, $\dot M_\bullet \propto L_{\rm x}$,
in a separate work, we will focus on the integrated properties of the proposed unified model, as the total black hole masses and related scalings (e.g., the Magorrian relation). We anticipate some important considerations.
As discussed above, the CCA self-regulation has a characteristic frequency related to the cooling time, $1/t_{\rm cool}$, as 
the hot halo requires such time to promote condensation, rain down, and then activate the ultra-fast outflow feedback. One loop requires $t_{\rm cyc} = t_{\rm cool} + t_{\rm OUT}\approx t_{\rm cool}$ (the outflow active time is always shorter than the condensation time).
In other words, the duty cycle increases from clusters to galaxies, as corroborated by long-term AGN feedback simulations (e.g., \citealt{Gaspari:2011a,Gaspari:2011b,Gaspari:2012b}) and X-ray shocks/cavities observations (e.g., \citealt{Randall:2015}). 
The number of cycles over the Hubble time is thus $n_{\rm cyc} = t_{\rm H}/t_{\rm cool}$, with an active time $t_{\rm act}= n_{\rm cyc}\,t_{\rm OUT}$. 
The black hole masses are expected to grow as $M_\bullet \simeq \dot M_\bullet\,t_{\rm act}$, hence with a temperature scaling given by $M_\bullet\propto L_{\rm x}/t_{\rm cool} \propto T_{\rm x}^3/(T_{\rm x}/\Lambda) \propto T_{\rm x}^2 \propto \sigma^4_\ast$, as core temperature is a measure of the (stellar) velocity variance in virialized structures. This is valid in the galactic regime ($T_{\rm x}\approx 0.5$\,-\,2\,keV) as $\Lambda$ remains essentially constant for solar metallicity. For clusters, $\Lambda \propto T^{1/2}$ due to Bremsstrahlung, thus $M_\bullet \propto T_{\rm x}^{2.5} \propto \sigma_\ast^{5}$. Observations show a very similar scaling, with ultramassive black holes found predominantly in more massive halos which are consistent with our self-regulated CCA model inducing a steepening of the Magorrian relation (e.g., \citealt{Gultekin:2009,McConnell:2013,Kormendy:2013}).

\section{Summary and Conclusions} \label{s:conc}
\noindent
We linked for the first time the {\it physical} micro and macro mechanical efficiency of SMBHs, the latter based on key X-ray data and hydrodynamical simulations, the former retrieved by state-of-the-art GR-RMHD horizon simulations, such that $\varepsilon_{\rm BH} = 10^{-3}\,T_{\rm x,7.4}$ and $\varepsilon_\bullet = 0.03$, respectively (\S\ref{s:efficiencies}).
By using minimally first principles, as conservation of energy ($P_{\rm out} = P_{\rm OUT} \simeq L_{\rm x}$, where the latter is the core luminosity of the hot halo), we unified the macro and micro properties of self-regulated AGN feedback from the galactic to the cluster regime (\S\ref{s:link}).

The inflow mechanism occurs via chaotic cold accretion (CCA) -- probed during the last years -- i.e., the rain of cold clouds condensing out of the quenched cooling flow ($\dot M_{\rm cool}$), which are recurrently funneled via fractal, inelastic collisions.
Near hundreds gravitational radii, the binding energy of accreting gas is strongly transformed into ultrafast outflows (UFOs) with characteristic velocity of a few $10^4$\,km\,s$^{-1}$ ($\sqrt{2\,\varepsilon_{\rm BH}}\,c$) 
ejecting most of the inflowing gas mass as $\dot M_{\rm out}\approx \dot M_{\rm cool}$ ($\approx 1\,\msun$\,yr$^{-1}$ for intermediate systems).

At larger radii, the outflow entrains progressively more mass, such as $\dot M_{\rm OUT}=\eta\,\dot M_{\rm out}$ and $v_{\rm OUT} = \eta^{-1/2}\,v_{\rm out}$, with $\eta \propto r^{2/3}$.
At roughly the kpc scale, the characteristic velocities of large-scale hot/warm/cold outflows are predicted to be a few $10^3$, 1000, and 500\,km\,s$^{-1}$, respectively (depending on the inner dominant gas phase). The related average mass outflow rates (for 1\,keV systems) are expected to be of the order of 10, 100, several 100\,$\msun$\,yr$^{-1}$, respectively. Such properties are in agreement with observations of UFOs, and kpc-scale ionized, neutral, and molecular outflows (\S\ref{s:comp}). Velocities are the more robust and stable indicator compared with outflow rates, both observationally and in the model.
Ultimately, the outflows thermalize within the system core ($\lta0.1\,r_{\rm vir}$), balancing the cooling losses, and allowing another self-regulated loop to reload via CCA rain and outflow feedback -- with frequency $\propto t_{\rm cool}^{-1}$.

A key aspect of the newly presented model is that the irradiated cool-core energy rate ($L_{\rm x}$) reflects the gas flow onto the tiny SMBH, creating a symbiotic link over a 10 dex dynamical range. The tiny SMBHs are not isolated point objects where space-time diverges, but appear to be central actors in the evolution of both the micro and cosmic structures. In particular, the SMBH growth rate is linked to the large-scale $T_{\rm x}$ halo and thus any other cosmic scaling (e.g., $L_{\rm x}, M_{\rm vir}$), in addition to inducing a consistent $M_\bullet - \sigma_\ast$ relation.
Despite the necessary limitations (\S\ref{s:disc}), the CCA$+$UFO model captures the essential ingredients than any more sophisticated self-regulation model and simulation should have at its core, in particular the gentle quasi-thermal equilibrium of plasma halos. 

The pursued minimalism of the CCA$+$UFO model makes it suited to be trivially implemented in subgrid modules and semi-analytic works (\S\ref{s:subgrid}), as well as in estimates for the interpretation of observational studies, e.g., related to nuclear and entrained outflow velocities and mass rates.
The proposed model presents a simple {\it physical} unification scheme upon which construct and conduct future multiwavelength investigations, e.g., selecting the systems in terms of the core X-ray luminosity (or other related macro  observable). 
Instead of classifying a phenomenological aspect of a peculiar AGN,
we encourage observational campaigns in the direction of understanding the common, unified physics of multiphase inflows/outflows (e.g., \S\ref{s:comp}) and to systematically consider the connection between the AGN and the global hot halo. A larger and homogeneous X-ray, optical, and radio sample of such properties, from low-mass galaxies to massive clusters, is needed to robustly test the link of the micro and macro properties of AGN feedback.

\section*{Acknowledgements}
\noindent
MG and AS acknowledge support for this work by NASA through Einstein Postdoctotral Fellowship number PF5-160137 and PF4-150126 awarded by the Chandra X-ray Center, which is operated by the Smithsonian Astrophysical Observatory for NASA under contract NAS8-03060. Support for this work was also provided by NASA Chandra award number G07-18121X.
FLASH code was in part developed by the DOE NNSA-ASC OASCR Flash center at the University of Chicago.
HPC resources were provided by the PL-Grid Infrastructure and the NASA/Ames HEC Program (SMD-16-7251).
We thank B. McNamara, G. Tremblay, J. Stone, M. McDonald, R. Morganti, F. Tombesi, M. Cappi, and F. Combes for insightful discussions.

\bibliographystyle{mnras}
\bibliography{biblio}

\providecommand{\SortNoop}[1]{}
\begin{thebibliography}{}
\makeatletter
\relax
\def\mn@urlcharsother{\let\do\@makeother \do\$\do\&\do\#\do\^\do\_\do\%\do\~}
\def\mn@doi{\begingroup\mn@urlcharsother \@ifnextchar [ {\mn@doi@}
  {\mn@doi@[]}}
\def\mn@doi@[#1]#2{\def\@tempa{#1}\ifx\@tempa\@empty \href
  {http://dx.doi.org/#2} {doi:#2}\else \href {http://dx.doi.org/#2} {#1}\fi
  \endgroup}
\def\mn@eprint#1#2{\mn@eprint@#1:#2::\@nil}
\def\mn@eprint@arXiv#1{\href {http://arxiv.org/abs/#1} {{\tt arXiv:#1}}}
\def\mn@eprint@dblp#1{\href {http://dblp.uni-trier.de/rec/bibtex/#1.xml}
  {dblp:#1}}
\def\mn@eprint@#1:#2:#3:#4\@nil{\def\@tempa {#1}\def\@tempb {#2}\def\@tempc
  {#3}\ifx \@tempc \@empty \let \@tempc \@tempb \let \@tempb \@tempa \fi \ifx
  \@tempb \@empty \def\@tempb {arXiv}\fi \@ifundefined
  {mn@eprint@\@tempb}{\@tempb:\@tempc}{\expandafter \expandafter \csname
  mn@eprint@\@tempb\endcsname \expandafter{\@tempc}}}

\bibitem[\protect\citeauthoryear{{Anderson}, {Gaspari}, {White}, {Wang}  \&
  {Dai}}{{Anderson} et~al.}{2015}]{Anderson:2015}
{Anderson} M.~E.,  {Gaspari} M.,  {White} S.~D.~M.,  {Wang} W.,   {Dai} X.,
  2015, \mn@doi [\mnras] {10.1093/mnras/stv437}, \href
  {http://adsabs.harvard.edu/abs/2015MNRAS.449.3806A} {449, 3806}

\bibitem[\protect\citeauthoryear{{Bondi}}{{Bondi}}{1952}]{Bondi:1952}
{Bondi} H.,  1952, \mnras, \href
  {http://adsabs.harvard.edu/abs/1952MNRAS.112..195B} {112, 195}

\bibitem[\protect\citeauthoryear{{Cicone} et~al.,}{{Cicone}
  et~al.}{2014}]{Cicone:2014}
{Cicone} C.,  et~al., 2014, \mn@doi [\aap] {10.1051/0004-6361/201322464}, \href
  {http://adsabs.harvard.edu/abs/2014A%26A...562A..21C} {562, A21}

\bibitem[\protect\citeauthoryear{{Combes}}{{Combes}}{2015}]{Combes:2015}
{Combes} F.,  2015, in {Ziegler} B.~L.,  {Combes} F.,  {Dannerbauer} H.,
  {Verdugo} M.,  eds,  IAU Symposium Vol. 309, Galaxies in 3D across the
  Universe. pp 182--189 (\mn@eprint {arXiv} {1408.1591}),
  \mn@doi{10.1017/S1743921314009636}

\bibitem[\protect\citeauthoryear{{Crenshaw}, {Kraemer}  \& {George}}{{Crenshaw}
  et~al.}{2003}]{Crenshaw:2003}
{Crenshaw} D.~M.,  {Kraemer} S.~B.,   {George} I.~M.,  2003, \mn@doi [\araa]
  {10.1146/annurev.astro.41.082801.100328}, \href
  {http://adsabs.harvard.edu/abs/2003ARA%26A..41..117C} {41, 117}

\bibitem[\protect\citeauthoryear{{David} et~al.,}{{David}
  et~al.}{2014}]{David:2014}
{David} L.~P.,  et~al., 2014, \mn@doi [\apj] {10.1088/0004-637X/792/2/94},
  \href {http://adsabs.harvard.edu/abs/2014ApJ...792...94D} {792, 94}

\bibitem[\protect\citeauthoryear{{Ettori} \& {Fabian}}{{Ettori} \&
  {Fabian}}{2000}]{Ettori:2000}
{Ettori} S.,  {Fabian} A.~C.,  2000, \mn@doi [\mnras]
  {10.1046/j.1365-8711.2000.03899.x}, \href
  {http://adsabs.harvard.edu/abs/2000MNRAS.317L..57E} {317, L57}

\bibitem[\protect\citeauthoryear{{Faucher-Gigu{\`e}re} \&
  {Quataert}}{{Faucher-Gigu{\`e}re} \& {Quataert}}{2012}]{Faucher:2012}
{Faucher-Gigu{\`e}re} C.-A.,  {Quataert} E.,  2012, \mn@doi [\mnras]
  {10.1111/j.1365-2966.2012.21512.x}, \href
  {http://adsabs.harvard.edu/abs/2012MNRAS.425..605F} {425, 605}

\bibitem[\protect\citeauthoryear{{Feruglio}, {Maiolino}, {Piconcelli}, {Menci},
  {Aussel}, {Lamastra}  \& {Fiore}}{{Feruglio} et~al.}{2010}]{Feruglio:2010}
{Feruglio} C.,  {Maiolino} R.,  {Piconcelli} E.,  {Menci} N.,  {Aussel} H.,
  {Lamastra} A.,   {Fiore} F.,  2010, \mn@doi [\aap]
  {10.1051/0004-6361/201015164}, \href
  {http://adsabs.harvard.edu/abs/2010A%26A...518L.155F} {518, L155}

\bibitem[\protect\citeauthoryear{{Feruglio} et~al.,}{{Feruglio}
  et~al.}{2015}]{Feruglio:2015}
{Feruglio} C.,  et~al., 2015, \mn@doi [\aap] {10.1051/0004-6361/201526020},
  \href {http://adsabs.harvard.edu/abs/2015A%26A...583A..99F} {583, A99}

\bibitem[\protect\citeauthoryear{{Fukumura}, {Kazanas}, {Contopoulos}  \&
  {Behar}}{{Fukumura} et~al.}{2010}]{Fukumura:2010}
{Fukumura} K.,  {Kazanas} D.,  {Contopoulos} I.,   {Behar} E.,  2010, \mn@doi
  [\apj] {10.1088/0004-637X/715/1/636}, \href
  {http://adsabs.harvard.edu/abs/2010ApJ...715..636F} {715, 636}

\bibitem[\protect\citeauthoryear{{Fukumura}, {Tombesi}, {Kazanas}, {Shrader},
  {Behar}  \& {Contopoulos}}{{Fukumura} et~al.}{2014}]{Fukumura:2014}
{Fukumura} K.,  {Tombesi} F.,  {Kazanas} D.,  {Shrader} C.,  {Behar} E.,
  {Contopoulos} I.,  2014, \mn@doi [\apj] {10.1088/0004-637X/780/2/120}, \href
  {http://adsabs.harvard.edu/abs/2014ApJ...780..120F} {780, 120}

\bibitem[\protect\citeauthoryear{{Fukumura}, {Tombesi}, {Kazanas}, {Shrader},
  {Behar}  \& {Contopoulos}}{{Fukumura} et~al.}{2015}]{Fukumura:2015}
{Fukumura} K.,  {Tombesi} F.,  {Kazanas} D.,  {Shrader} C.,  {Behar} E.,
  {Contopoulos} I.,  2015, \mn@doi [\apj] {10.1088/0004-637X/805/1/17}, \href
  {http://adsabs.harvard.edu/abs/2015ApJ...805...17F} {805, 17}

\bibitem[\protect\citeauthoryear{{Gaspari}}{{Gaspari}}{2015}]{Gaspari:2015_xspec}
{Gaspari} M.,  2015, \mn@doi [\mnras] {10.1093/mnrasl/slv067}, \href
  {http://adsabs.harvard.edu/abs/2015MNRAS.451L..60G} {451, L60}

\bibitem[\protect\citeauthoryear{{Gaspari}, {{\SortNoop{m2011a}}Melioli},
  {Brighenti}  \& {D'Ercole}}{{Gaspari} et~al.}{2011a}]{Gaspari:2011a}
{Gaspari} M.,  {{\SortNoop{m2011a}}Melioli} C.,  {Brighenti} F.,   {D'Ercole}
  A.,  2011a, \mn@doi [\mnras] {10.1111/j.1365-2966.2010.17688.x}, \href
  {http://adsabs.harvard.edu/abs/2011MNRAS.411..349G} {411, 349}

\bibitem[\protect\citeauthoryear{{Gaspari}, {{\SortNoop{m2011b}}Brighenti},
  {D'Ercole}  \& {Melioli}}{{Gaspari} et~al.}{2011b}]{Gaspari:2011b}
{Gaspari} M.,  {{\SortNoop{m2011b}}Brighenti} F.,  {D'Ercole} A.,   {Melioli}
  C.,  2011b, \mn@doi [\mnras] {10.1111/j.1365-2966.2011.18806.x}, \href
  {http://adsabs.harvard.edu/abs/2011MNRAS.415.1549G} {415, 1549}

\bibitem[\protect\citeauthoryear{{Gaspari}, {\SortNoop{b}}{Brighenti}  \&
  {Temi}}{{Gaspari} et~al.}{2012a}]{Gaspari:2012b}
{Gaspari} M.,  {\SortNoop{b}}{Brighenti} F.,   {Temi} P.,  2012a, \mn@doi
  [\mnras] {10.1111/j.1365-2966.2012.21183.x}, \href
  {http://adsabs.harvard.edu/abs/2012MNRAS.424..190G} {424, 190}

\bibitem[\protect\citeauthoryear{{Gaspari}, {\SortNoop{a}}{Ruszkowski}  \&
  {Sharma}}{{Gaspari} et~al.}{2012b}]{Gaspari:2012a}
{Gaspari} M.,  {\SortNoop{a}}{Ruszkowski} M.,   {Sharma} P.,  2012b, \mn@doi
  [\apj] {10.1088/0004-637X/746/1/94}, \href
  {http://adsabs.harvard.edu/abs/2012ApJ...746...94G} {746, 94}

\bibitem[\protect\citeauthoryear{{Gaspari}, {Ruszkowski}  \& {Oh}}{{Gaspari}
  et~al.}{2013}]{Gaspari:2013_cca}
{Gaspari} M.,  {Ruszkowski} M.,   {Oh} S.~P.,  2013, \mn@doi [\mnras]
  {10.1093/mnras/stt692}, \href
  {http://adsabs.harvard.edu/abs/2013MNRAS.432.3401G} {432, 3401}

\bibitem[\protect\citeauthoryear{{Gaspari}, {Brighenti}, {Temi}  \&
  {Ettori}}{{Gaspari} et~al.}{2014}]{Gaspari:2014_scalings}
{Gaspari} M.,  {Brighenti} F.,  {Temi} P.,   {Ettori} S.,  2014, \mn@doi
  [\apjl] {10.1088/2041-8205/783/1/L10}, \href
  {http://adsabs.harvard.edu/abs/2014ApJ...783L..10G} {783, L10}

\bibitem[\protect\citeauthoryear{{Gaspari}, {Brighenti}  \& {Temi}}{{Gaspari}
  et~al.}{2015}]{Gaspari:2015_cca}
{Gaspari} M.,  {Brighenti} F.,   {Temi} P.,  2015, \mn@doi [\aap]
  {10.1051/0004-6361/201526151}, \href
  {http://adsabs.harvard.edu/abs/2015A%26A...579A..62G} {579, A62}

\bibitem[\protect\citeauthoryear{{Gaspari}, {Temi}  \& {Brighenti}}{{Gaspari}
  et~al.}{2017}]{Gaspari:2017}
{Gaspari} M.,  {Temi} P.,   {Brighenti} F.,  2017, \mn@doi [\mnras]
  {10.1093/mnras/stw3108}, \href
  {http://adsabs.harvard.edu/abs/2017MNRAS.466..677G} {466, 677}

\bibitem[\protect\citeauthoryear{{G{\"u}ltekin} et~al.,}{{G{\"u}ltekin}
  et~al.}{2009}]{Gultekin:2009}
{G{\"u}ltekin} K.,  et~al., 2009, \mn@doi [\apj] {10.1088/0004-637X/698/1/198},
  \href {http://adsabs.harvard.edu/abs/2009ApJ...698..198G} {698, 198}

\bibitem[\protect\citeauthoryear{{Heckman} \& {Best}}{{Heckman} \&
  {Best}}{2014}]{Heckman:2014}
{Heckman} T.~M.,  {Best} P.~N.,  2014, \mn@doi [\araa]
  {10.1146/annurev-astro-081913-035722}, \href
  {http://adsabs.harvard.edu/abs/2014ARA%26A..52..589H} {52, 589}

\bibitem[\protect\citeauthoryear{{Kaastra} et~al.,}{{Kaastra}
  et~al.}{2004}]{Kaastra:2004}
{Kaastra} J.~S.,  et~al., 2004, \mn@doi [\aap] {10.1051/0004-6361:20031512},
  \href {http://adsabs.harvard.edu/abs/2004A%26A...413..415K} {413, 415}

\bibitem[\protect\citeauthoryear{{Khatri} \& {Gaspari}}{{Khatri} \&
  {Gaspari}}{2016}]{Khatri:2016}
{Khatri} R.,  {Gaspari} M.,  2016, \mn@doi [\mnras] {10.1093/mnras/stw2027},
  \href {http://adsabs.harvard.edu/abs/2016MNRAS.463..655K} {463, 655}

\bibitem[\protect\citeauthoryear{{King} \& {Nixon}}{{King} \&
  {Nixon}}{2015}]{King:2015_flicker}
{King} A.,  {Nixon} C.,  2015, \mn@doi [\mnras] {10.1093/mnrasl/slv098}, \href
  {http://adsabs.harvard.edu/abs/2015MNRAS.453L..46K} {453, L46}

\bibitem[\protect\citeauthoryear{{King} \& {Pounds}}{{King} \&
  {Pounds}}{2014}]{King:2014_WA}
{King} A.~R.,  {Pounds} K.~A.,  2014, \mn@doi [\mnras] {10.1093/mnrasl/slt144},
  \href {http://adsabs.harvard.edu/abs/2014MNRAS.437L..81K} {437, L81}

\bibitem[\protect\citeauthoryear{{King} \& {Pringle}}{{King} \&
  {Pringle}}{2006}]{King:2006}
{King} A.~R.,  {Pringle} J.~E.,  2006, \mn@doi [\mnras]
  {10.1111/j.1745-3933.2006.00249.x}, \href
  {http://adsabs.harvard.edu/abs/2006MNRAS.373L..90K} {373, L90}

\bibitem[\protect\citeauthoryear{{Kormendy} \& {Ho}}{{Kormendy} \&
  {Ho}}{2013}]{Kormendy:2013}
{Kormendy} J.,  {Ho} L.~C.,  2013, \mn@doi [\araa]
  {10.1146/annurev-astro-082708-101811}, \href
  {http://adsabs.harvard.edu/abs/2013ARA%26A..51..511K} {51, 511}

\bibitem[\protect\citeauthoryear{{Kravtsov} \& {Borgani}}{{Kravtsov} \&
  {Borgani}}{2012}]{Kravtsov:2012}
{Kravtsov} A.~V.,  {Borgani} S.,  2012, \mn@doi [\araa]
  {10.1146/annurev-astro-081811-125502}, \href
  {http://adsabs.harvard.edu/abs/2012ARA%26A..50..353K} {50, 353}

\bibitem[\protect\citeauthoryear{{Main}, {McNamara}, {Nulsen}, {Russell}  \&
  {Vantyghem}}{{Main} et~al.}{2017}]{Main:2017}
{Main} R.~A.,  {McNamara} B.~R.,  {Nulsen} P.~E.~J.,  {Russell} H.~R.,
  {Vantyghem} A.~N.,  2017, \mn@doi [\mnras] {10.1093/mnras/stw2644}, \href
  {http://adsabs.harvard.edu/abs/2017MNRAS.464.4360M} {464, 4360}

\bibitem[\protect\citeauthoryear{{McConnell} \& {Ma}}{{McConnell} \&
  {Ma}}{2013}]{McConnell:2013}
{McConnell} N.~J.,  {Ma} C.-P.,  2013, \mn@doi [\apj]
  {10.1088/0004-637X/764/2/184}, \href
  {http://adsabs.harvard.edu/abs/2013ApJ...764..184M} {764, 184}

\bibitem[\protect\citeauthoryear{{McDonald}, {Roediger}, {Veilleux}  \&
  {Ehlert}}{{McDonald} et~al.}{2014}]{McDonald:2014}
{McDonald} M.,  {Roediger} J.,  {Veilleux} S.,   {Ehlert} S.,  2014, \mn@doi
  [\apjl] {10.1088/2041-8205/791/2/L30}, \href
  {http://adsabs.harvard.edu/abs/2014ApJ...791L..30M} {791, L30}

\bibitem[\protect\citeauthoryear{{McDonald} et~al.,}{{McDonald}
  et~al.}{2016}]{McDonald:2016}
{McDonald} M.,  et~al., 2016, \mn@doi [\apj] {10.3847/0004-637X/826/2/124},
  \href {http://adsabs.harvard.edu/abs/2016ApJ...826..124M} {826, 124}

\bibitem[\protect\citeauthoryear{{McDonald} et~al.,}{{McDonald}
  et~al.}{2017}]{McDonald:2017}
{McDonald} M.,  et~al., 2017, preprint, \href
  {http://adsabs.harvard.edu/abs/2017arXiv170205094M} {} (\mn@eprint {arXiv}
  {1702.05094})

\bibitem[\protect\citeauthoryear{{McNamara} \& {Nulsen}}{{McNamara} \&
  {Nulsen}}{2012}]{McNamara:2012}
{McNamara} B.~R.,  {Nulsen} P.~E.~J.,  2012, \mn@doi [New J. Phys.]
  {10.1088/1367-2630/14/5/055023}, \href
  {http://adsabs.harvard.edu/abs/2012NJPh...14e5023M} {14, 055023}

\bibitem[\protect\citeauthoryear{{McNamara}, {Russell}, {Nulsen}, {Hogan},
  {Fabian}, {Pulido}  \& {Edge}}{{McNamara} et~al.}{2016}]{McNamara:2016}
{McNamara} B.~R.,  {Russell} H.~R.,  {Nulsen} P.~E.~J.,  {Hogan} M.~T.,
  {Fabian} A.~C.,  {Pulido} F.,   {Edge} A.~C.,  2016, \mn@doi [\apj]
  {10.3847/0004-637X/830/2/79}, \href
  {http://adsabs.harvard.edu/abs/2016ApJ...830...79M} {830, 79}

\bibitem[\protect\citeauthoryear{{Merloni} \& {Heinz}}{{Merloni} \&
  {Heinz}}{2008}]{Merloni:2008}
{Merloni} A.,  {Heinz} S.,  2008, \mn@doi [\mnras]
  {10.1111/j.1365-2966.2008.13472.x}, \href
  {http://adsabs.harvard.edu/abs/2008MNRAS.388.1011M} {388, 1011}

\bibitem[\protect\citeauthoryear{{Moller} \& {Sadowski}}{{Moller} \&
  {Sadowski}}{2015}]{Moller:2015}
{Moller} A.,  {Sadowski} A.,  2015, preprint, \href
  {http://adsabs.harvard.edu/abs/2015arXiv150906644M} {} (\mn@eprint {arXiv}
  {1509.06644})

\bibitem[\protect\citeauthoryear{{Morganti}}{{Morganti}}{2015}]{Morganti:2015_rev}
{Morganti} R.,  2015, in {Massaro} F.,  {Cheung} C.~C.,  {Lopez} E.,
  {Siemiginowska} A.,  eds,  IAU Symposium Vol. 313, Extragalactic Jets from
  Every Angle. pp 283--288 (\mn@eprint {arXiv} {1411.6107}),
  \mn@doi{10.1017/S1743921315002331}

\bibitem[\protect\citeauthoryear{{Morganti}, {Tadhunter}  \&
  {Oosterloo}}{{Morganti} et~al.}{2005}]{Morganti:2005}
{Morganti} R.,  {Tadhunter} C.~N.,   {Oosterloo} T.~A.,  2005, \mn@doi [\aap]
  {10.1051/0004-6361:200500197}, \href
  {http://adsabs.harvard.edu/abs/2005A%26A...444L...9M} {444, L9}

\bibitem[\protect\citeauthoryear{{Morganti}, {Holt}, {Saripalli}, {Oosterloo}
  \& {Tadhunter}}{{Morganti} et~al.}{2007}]{Morganti:2007}
{Morganti} R.,  {Holt} J.,  {Saripalli} L.,  {Oosterloo} T.~A.,   {Tadhunter}
  C.~N.,  2007, \mn@doi [\aap] {10.1051/0004-6361:20077888}, \href
  {http://adsabs.harvard.edu/abs/2007A%26A...476..735M} {476, 735}

\bibitem[\protect\citeauthoryear{{Morganti}, {Oosterloo}, {Oonk}, {Frieswijk}
  \& {Tadhunter}}{{Morganti} et~al.}{2015}]{Morganti:2015_ALMA}
{Morganti} R.,  {Oosterloo} T.,  {Oonk} J.~B.~R.,  {Frieswijk} W.,
  {Tadhunter} C.,  2015, \mn@doi [\aap] {10.1051/0004-6361/201525860}, \href
  {http://adsabs.harvard.edu/abs/2015A%26A...580A...1M} {580, A1}

\bibitem[\protect\citeauthoryear{{Narayan} \& {Yi}}{{Narayan} \&
  {Yi}}{1995}]{Narayan:1995}
{Narayan} R.,  {Yi} I.,  1995, \mn@doi [\apj] {10.1086/176343}, \href
  {http://adsabs.harvard.edu/abs/1995ApJ...452..710N} {452, 710}

\bibitem[\protect\citeauthoryear{{Nesvadba} et~al.,}{{Nesvadba}
  et~al.}{2010}]{Nesvadba:2010}
{Nesvadba} N.~P.~H.,  et~al., 2010, \mn@doi [\aap]
  {10.1051/0004-6361/200913333}, \href
  {http://adsabs.harvard.edu/abs/2010A%26A...521A..65N} {521, A65}

\bibitem[\protect\citeauthoryear{{Novikov} \& {Thorne}}{{Novikov} \&
  {Thorne}}{1973}]{Novikov:1973}
{Novikov} I.~D.,  {Thorne} K.~S.,  1973, in {Dewitt} C.,  {Dewitt} B.~S.,  eds,
  Black Holes (Les Astres Occlus). pp 343--450

\bibitem[\protect\citeauthoryear{{Peterson}, {Kahn}, {Paerels}, {Kaastra},
  {Tamura}, {Bleeker}, {Ferrigno}  \& {Jernigan}}{{Peterson}
  et~al.}{2003}]{Peterson:2003}
{Peterson} J.~R.,  {Kahn} S.~M.,  {Paerels} F.~B.~S.,  {Kaastra} J.~S.,
  {Tamura} T.,  {Bleeker} J.~M.,  {Ferrigno} C.,   {Jernigan} J.,  2003,
  \mn@doi [\apj] {10.1086/374830}, \href
  {http://adsabs.harvard.edu/abs/2003ApJ...590..207P} {590, 207}

\bibitem[\protect\citeauthoryear{{Pizzolato} \& {Soker}}{{Pizzolato} \&
  {Soker}}{2010}]{Pizzolato:2010}
{Pizzolato} F.,  {Soker} N.,  2010, \mn@doi [\mnras]
  {10.1111/j.1365-2966.2010.17156.x}, \href
  {http://adsabs.harvard.edu/abs/2010MNRAS.408..961P} {408, 961}

\bibitem[\protect\citeauthoryear{{Prasad}, {Sharma}  \& {Babul}}{{Prasad}
  et~al.}{2015}]{Prasad:2015}
{Prasad} D.,  {Sharma} P.,   {Babul} A.,  2015, \mn@doi [\apj]
  {10.1088/0004-637X/811/2/108}, \href
  {http://adsabs.harvard.edu/abs/2015ApJ...811..108P} {811, 108}

\bibitem[\protect\citeauthoryear{{Randall} et~al.,}{{Randall}
  et~al.}{2015}]{Randall:2015}
{Randall} S.~W.,  et~al., 2015, \mn@doi [\apj] {10.1088/0004-637X/805/2/112},
  \href {http://adsabs.harvard.edu/abs/2015ApJ...805..112R} {805, 112}

\bibitem[\protect\citeauthoryear{{Russell}, {McNamara}, {Edge}, {Hogan}, {Main}
   \& {Vantyghem}}{{Russell} et~al.}{2013}]{Russell:2013}
{Russell} H.~R.,  {McNamara} B.~R.,  {Edge} A.~C.,  {Hogan} M.~T.,  {Main}
  R.~A.,   {Vantyghem} A.~N.,  2013, \mn@doi [\mnras] {10.1093/mnras/stt490},
  \href {http://adsabs.harvard.edu/abs/2013MNRAS.432..530R} {432, 530}

\bibitem[\protect\citeauthoryear{{Russell} et~al.,}{{Russell}
  et~al.}{2014}]{Russell:2014}
{Russell} H.~R.,  et~al., 2014, \mn@doi [\apj] {10.1088/0004-637X/784/1/78},
  \href {http://adsabs.harvard.edu/abs/2014ApJ...784...78R} {784, 78}

\bibitem[\protect\citeauthoryear{{Russell} et~al.,}{{Russell}
  et~al.}{2016}]{Russell:2016_phoenix}
{Russell} H.~R.,  et~al., 2016, preprint, \href
  {http://adsabs.harvard.edu/abs/2016arXiv161100017R} {} (\mn@eprint {arXiv}
  {1611.00017})

\bibitem[\protect\citeauthoryear{{Sadowski} \& {Gaspari}}{{Sadowski} \&
  {Gaspari}}{2017}]{Sadowski:2017}
{Sadowski} A.,  {Gaspari} M.,  2017, preprint, \href
  {http://adsabs.harvard.edu/abs/2017arXiv170107033S} {} (\mn@eprint {arXiv}
  {1701.07033})

\bibitem[\protect\citeauthoryear{{S{\c a}dowski}, {Narayan}, {Tchekhovskoy},
  {Abarca}, {Zhu}  \& {McKinney}}{{S{\c a}dowski} et~al.}{2015}]{Sadowski:2015}
{S{\c a}dowski} A.,  {Narayan} R.,  {Tchekhovskoy} A.,  {Abarca} D.,  {Zhu} Y.,
    {McKinney} J.~C.,  2015, \mn@doi [\mnras] {10.1093/mnras/stu2387}, \href
  {http://adsabs.harvard.edu/abs/2015MNRAS.447...49S} {447, 49}

\bibitem[\protect\citeauthoryear{{S{\c a}dowski}, {Lasota}, {Abramowicz}  \&
  {Narayan}}{{S{\c a}dowski} et~al.}{2016}]{Sadowski:2016_thick}
{S{\c a}dowski} A.,  {Lasota} J.-P.,  {Abramowicz} M.~A.,   {Narayan} R.,
  2016, \mn@doi [\mnras] {10.1093/mnras/stv2854}, \href
  {http://adsabs.harvard.edu/abs/2016MNRAS.456.3915S} {456, 3915}

\bibitem[\protect\citeauthoryear{{Soker}}{{Soker}}{2016}]{Soker:2016}
{Soker} N.,  2016, \mn@doi [\nar] {10.1016/j.newar.2016.08.002}, \href
  {http://adsabs.harvard.edu/abs/2016NewAR..75....1S} {75, 1}

\bibitem[\protect\citeauthoryear{{Soker}, {Sternberg}  \& {Pizzolato}}{{Soker}
  et~al.}{2009}]{Soker:2009}
{Soker} N.,  {Sternberg} A.,   {Pizzolato} F.,  2009, in {Heinz} S.,  {Wilcots}
  E.,  eds,  American Institute of Physics Conference Series Vol. 1201,
  American Institute of Physics Conference Series. pp 321--325 (\mn@eprint
  {arXiv} {0909.0220}), \mn@doi{10.1063/1.3293066}

\bibitem[\protect\citeauthoryear{{Sturm} et~al.,}{{Sturm}
  et~al.}{2011}]{Sturm:2011}
{Sturm} E.,  et~al., 2011, \mn@doi [\apjl] {10.1088/2041-8205/733/1/L16}, \href
  {http://adsabs.harvard.edu/abs/2011ApJ...733L..16S} {733, L16}

\bibitem[\protect\citeauthoryear{{Sun}}{{Sun}}{2012}]{Sun:2012}
{Sun} M.,  2012, \mn@doi [New Journal of Physics]
  {10.1088/1367-2630/14/4/045004}, \href
  {http://adsabs.harvard.edu/abs/2012NJPh...14d5004S} {14, 045004}

\bibitem[\protect\citeauthoryear{{Sun}, {Voit}, {Donahue}, {Jones}, {Forman}
  \& {Vikhlinin}}{{Sun} et~al.}{2009}]{Sun:2009a}
{Sun} M.,  {Voit} G.~M.,  {Donahue} M.,  {Jones} C.,  {Forman} W.,
  {Vikhlinin} A.,  2009, \mn@doi [\apj] {10.1088/0004-637X/693/2/1142}, \href
  {http://adsabs.harvard.edu/abs/2009ApJ...693.1142S} {693, 1142}

\bibitem[\protect\citeauthoryear{{Sutherland} \& {Dopita}}{{Sutherland} \&
  {Dopita}}{1993}]{Sutherland:1993}
{Sutherland} R.~S.,  {Dopita} M.~A.,  1993, \mn@doi [\apjs] {10.1086/191823},
  \href {http://adsabs.harvard.edu/abs/1993ApJS...88..253S} {88, 253}

\bibitem[\protect\citeauthoryear{{Tchekhovskoy}, {Narayan}  \&
  {McKinney}}{{Tchekhovskoy} et~al.}{2011}]{Tchekhovskoy:2011}
{Tchekhovskoy} A.,  {Narayan} R.,   {McKinney} J.~C.,  2011, \mn@doi [\mnras]
  {10.1111/j.1745-3933.2011.01147.x}, \href
  {http://adsabs.harvard.edu/abs/2011MNRAS.418L..79T} {418, L79}

\bibitem[\protect\citeauthoryear{{Teng}, {Veilleux}  \& {Baker}}{{Teng}
  et~al.}{2013}]{Teng:2013}
{Teng} S.~H.,  {Veilleux} S.,   {Baker} A.~J.,  2013, \mn@doi [\apj]
  {10.1088/0004-637X/765/2/95}, \href
  {http://adsabs.harvard.edu/abs/2013ApJ...765...95T} {765, 95}

\bibitem[\protect\citeauthoryear{{Tombesi}}{{Tombesi}}{2016}]{Tombesi:2016}
{Tombesi} F.,  2016, \mn@doi [Astronomische Nachrichten]
  {10.1002/asna.201612322}, \href
  {http://adsabs.harvard.edu/abs/2016AN....337..410T} {337, 410}

\bibitem[\protect\citeauthoryear{{Tombesi}, {Cappi}, {Reeves}  \&
  {Braito}}{{Tombesi} et~al.}{2012}]{Tombesi:2012}
{Tombesi} F.,  {Cappi} M.,  {Reeves} J.~N.,   {Braito} V.,  2012, \mn@doi
  [\mnras] {10.1111/j.1745-3933.2012.01221.x}, \href
  {http://adsabs.harvard.edu/abs/2012MNRAS.tmpL.413T} {p.~L413}

\bibitem[\protect\citeauthoryear{{Tombesi}, {Cappi}, {Reeves}, {Nemmen},
  {Braito}, {Gaspari}  \& {Reynolds}}{{Tombesi} et~al.}{2013}]{Tombesi:2013}
{Tombesi} F.,  {Cappi} M.,  {Reeves} J.~N.,  {Nemmen} R.~S.,  {Braito} V.,
  {Gaspari} M.,   {Reynolds} C.~S.,  2013, \mn@doi [\mnras]
  {10.1093/mnras/sts692}, \href
  {http://adsabs.harvard.edu/abs/2013MNRAS.430.1102T} {430, 1102}

\bibitem[\protect\citeauthoryear{{Tombesi}, {Tazaki}, {Mushotzky}, {Ueda},
  {Cappi}, {Gofford}, {Reeves}  \& {Guainazzi}}{{Tombesi}
  et~al.}{2014}]{Tombesi:2014}
{Tombesi} F.,  {Tazaki} F.,  {Mushotzky} R.~F.,  {Ueda} Y.,  {Cappi} M.,
  {Gofford} J.,  {Reeves} J.~N.,   {Guainazzi} M.,  2014, \mn@doi [\mnras]
  {10.1093/mnras/stu1297}, \href
  {http://adsabs.harvard.edu/abs/2014MNRAS.443.2154T} {443, 2154}

\bibitem[\protect\citeauthoryear{{Tombesi}, {Mel{\'e}ndez}, {Veilleux},
  {Reeves}, {Gonz{\'a}lez-Alfonso}  \& {Reynolds}}{{Tombesi}
  et~al.}{2015}]{Tombesi:2015}
{Tombesi} F.,  {Mel{\'e}ndez} M.,  {Veilleux} S.,  {Reeves} J.~N.,
  {Gonz{\'a}lez-Alfonso} E.,   {Reynolds} C.~S.,  2015, \mn@doi [\nat]
  {10.1038/nature14261}, \href
  {http://adsabs.harvard.edu/abs/2015Natur.519..436T} {519, 436}

\bibitem[\protect\citeauthoryear{{Tremblay} et~al.,}{{Tremblay}
  et~al.}{2016}]{Tremblay:2016}
{Tremblay} G.~R.,  et~al., 2016, \mn@doi [\nat] {10.1038/nature17969}, \href
  {http://adsabs.harvard.edu/abs/2016Natur.534..218T} {534, 218}

\bibitem[\protect\citeauthoryear{{Vikhlinin}, {Kravtsov}, {Forman}, {Jones},
  {Markevitch}, {Murray}  \& {Van Speybroeck}}{{Vikhlinin}
  et~al.}{2006}]{Vikhlinin:2006}
{Vikhlinin} A.,  {Kravtsov} A.,  {Forman} W.,  {Jones} C.,  {Markevitch} M.,
  {Murray} S.~S.,   {Van Speybroeck} L.,  2006, \mn@doi [\apj]
  {10.1086/500288}, \href {http://adsabs.harvard.edu/abs/2006ApJ...640..691V}
  {640, 691}

\bibitem[\protect\citeauthoryear{{Voit}, {Donahue}, {Bryan}  \&
  {McDonald}}{{Voit} et~al.}{2015}]{Voit:2015_nat}
{Voit} G.~M.,  {Donahue} M.,  {Bryan} G.~L.,   {McDonald} M.,  2015, \mn@doi
  [\nat] {10.1038/nature14167}, \href
  {http://adsabs.harvard.edu/abs/2015Natur.519..203V} {519, 203}

\bibitem[\protect\citeauthoryear{{Werner} et~al.,}{{Werner}
  et~al.}{2014}]{Werner:2014}
{Werner} N.,  et~al., 2014, \mn@doi [\mnras] {10.1093/mnras/stu006}, \href
  {http://adsabs.harvard.edu/abs/2014MNRAS.439.2291W} {439, 2291}

\bibitem[\protect\citeauthoryear{{Yang} \& {Reynolds}}{{Yang} \&
  {Reynolds}}{2016}]{Yang:2016}
{Yang} H.-Y.~K.,  {Reynolds} C.~S.,  2016, \mn@doi [\apj]
  {10.3847/0004-637X/829/2/90}, \href
  {http://adsabs.harvard.edu/abs/2016ApJ...829...90Y} {829, 90}

\bibitem[\protect\citeauthoryear{{Zubovas} \& {King}}{{Zubovas} \&
  {King}}{2014}]{Zubovas:2014}
{Zubovas} K.,  {King} A.~R.,  2014, \mn@doi [\mnras] {10.1093/mnras/stt2472},
  \href {http://adsabs.harvard.edu/abs/2014MNRAS.439..400Z} {439, 400}

\makeatother
\end{thebibliography}

\appendix
\section{Core luminosity and temperature}\label{app:L0}
\noindent
Most of the X-ray luminosity comes from the region well within $r_{500}$ 
due to the steep radial density profile (emissivity is $\propto \rho(r)^2$). 
By using the available {\it Chandra} and {\it XMM} (losing sensitivity at large radii) luminosities is thus a fair proxy for the core luminosity.
More accurately, we can model the surface brightness 
with a $\beta$ profile, ${\rm SB_x}={\rm SB_0}\,(1+R^2/r^2_{\rm c})^{-3\beta+1/2}$, where $R$ is the projected radius and ${\rm SB_0}$ is the inner normalization. Integrating over thin annuli yields
\begin{equation}\label{e:Lxbeta}
L_{\rm x}(<r) = {\rm SB_0}\,\frac{2\pi\,r^2_{\rm c}}{3(2\beta-1)}\left[1-\left(1+\frac{R^2}{r^2_{\rm c}}\right)^{-3\beta+3/2} \right].
\end{equation}
The cooling radius is typically equal to the core radius ($\approx 0.2\,r_{500}$; \citealt{Vikhlinin:2006}), since the radial breaking naturally emerges via the loss of pressure, $r_{\rm cool}\simeq r_{\rm c}$ (\citealt{Ettori:2000}). Cool-core systems are better fitted by a sum of two beta models for the core and the outskirt; characteristic values are $\beta_c\approx1.7$ and $\beta_o\approx0.7$, respectively (e.g., \citealt{Ettori:2000}).
Plugging in this values in a double $\beta$ model following each Eq.~\ref{e:Lxbeta}, the average correction for the core luminosity is $0.68\,L_{500}$. Notice that the outskirts, $r_{\rm vir}\simeq 2\,r_{500}$, contribute in negligible measure, $L_{\rm vir}/L_{\rm 500}\approx1.05$.
Overall, the chosen luminosity radius does not significantly alter the results presented in \S\ref{s:link}. 
The temperature profile shows even less variation than density, varying by a factor 2\,-\,3. By emission-weighting it, the core $T_{\rm x}$ is typically 10 percent lower than the ambient $T_{\rm 500}$ (\citealt{Ettori:2000,Vikhlinin:2006}) -- again, a minor variation.

For an idealized self-similar spherical collapse, it is well known that $L_{\rm x}\propto T_{\rm x}^2$ (\citealt{Kravtsov:2012}). However, observational data show that non-gravitational/feedback processes steepen such relation as $L_{\rm 500}\simeq 8.8\times10^{43}\,(T_{\rm 500}/{\rm 2.5\,keV})^3$ (\citealt{Sun:2012}). In \S\ref{s:link}, we are interested in the X-ray luminosity and temperature tied to the core/cooling region, i.e.,  the radius within which the temperature profile slope becomes positive
($r_{\rm c}\approx 0.2\,r_{500}$, related to $t_{\rm cool}\approx t_{\rm age}\sim t_{\rm H}/2$, where $t_{\rm H}$ is the Hubble time).
By using the above minor corrections, the {\it core} scaling relation becomes $L_{\rm x}\simeq 6\times10^{43}\,(T_{\rm x}/{\rm 2.2\,keV})^3$ erg\,s$^{-1}$.
For reference, in the local universe, the scaling between radius and temperature ($r^3\propto M\propto T^{3/2}$) is $r_{500}\simeq(0.74\,{\rm Mpc})\,T_{\rm x, 7.4}^{1/2}$ (\citealt{Sun:2009a}), leading to a physical core radius $r_{\rm c}\approx (148\,{\rm kpc})\,T_{\rm x,7.4}^{1/2}$.

\label{lastpage}
\end{document}